\documentclass[12pt]{article}

\usepackage[T1]{fontenc}
\usepackage[utf8]{inputenc}
\usepackage{float}
\usepackage{biblatex}
\usepackage{amsmath}
\usepackage{graphicx}
\usepackage{amsmath}
\usepackage{amssymb}
\usepackage{framed}
\usepackage{subfigure}
\usepackage[margin=16pt,font=small,labelfont=bf]{caption}
\usepackage[italian, english]{babel}
\usepackage{enumitem}
\usepackage{longtable}
\usepackage{mdwlist}
\usepackage{xcolor}
\usepackage{mdframed}
\usepackage[a4paper]{geometry}
\usepackage{tabularx}
\usepackage{booktabs}
\usepackage{multirow}
\usepackage{lscape}
\usepackage{wasysym}
\usepackage{tikz}
\usetikzlibrary{math}
\usepackage{pgfplots}
\usetikzlibrary{shapes}
\usetikzlibrary{arrows,automata}

\usepackage[colorlinks=true, linkcolor=black, citecolor=black, linktocpage=true]{hyperref}

\usepackage{parskip}

\newcommand{\qed}{\hspace*{\fill} $\bigstar$\medskip}

\def \Z {\mathbb Z}

\def \N {\mathbb N}

\def \cX {\mathcal X}

\def \epsi {\varepsilon}
\def \prm 	 {^{\prime}}

\def \tmix {T_{\text{mix}}}

\newcommand{\tvar}[2]{\|#1 - #2\|_{\text{TV}}}

\newcommand{\rightField}[2]{\overrightarrow{h_{#1}}({#2})}
\newcommand{\leftField}[2]{\overleftarrow{h_{#1}}({#2})}
\newcommand{\spinUp}[2]{{#2}_{{#1}^{\uparrow}}}
\newcommand{\spinRight}[2]{{#2}_{{#1}^{\rightarrow}}}
\newcommand{\spinDown}[2]{{#2}_{{#1}^{\downarrow}}}
\newcommand{\spinLeft}[2]{{#2}_{{#1}^{\leftarrow}}}
\newcommand{\spin}[2]{{#2}_{#1}}

\def \a {\alpha}

\pgfplotsset{
        compat=1.11,
        }
\newcommand{\myGlobalTransformation}[2]
{
    \pgftransformcm{1}{0}{0.4}{0.5}{\pgfpoint{#1cm}{#2cm}}
}

\newcommand{\gridThreeD}[3]
{
    \begin{scope}
        \myGlobalTransformation{#1}{#2};
        \draw [#3,step=2cm,xshift=1cm,yshift=1cm] grid (6,6);
    \end{scope}
}

\newcommand{\graphThreeDnodes}[2]
{
    \begin{scope}
        \myGlobalTransformation{#1}{#2};
        \foreach \x in {1,3,5,7} {
            \foreach \y in {1,3,5,7} {
                \node at (\x,\y) [circle,fill={black!50}, scale=0.3] {};
            }
        }
    \end{scope}
}
\tikzstyle myBG=[line width=0.2pt,opacity=1.0]
\pgfkeys{/pgf/declare function={arctanh(\x) = 0.5*(ln((1+\x)/(1-\x)));}}

\usepackage{algorithm}
\usepackage{algorithmic}

\author{
Roberto D'Autilia\\
Dipartimento di Matematica e Fisica\\
Universit\`a degli Studi Roma Tre\\
\and
Louis Nantenaina Andrianaivo\\
Dipartimento di Matematica e Fisica\\
Universit\`a degli Studi Roma Tre
\and
Alessio Troiani\\
Dipartimento di Matematica ``Tullio Levi-Civita''\\
Universit\`a degli Studi di Padova
}
\title{Parallel simulation of two--dimensional Ising models using Probabilistic Cellular Automata}
\date{\today}
\begin{document}

\maketitle

\begin{abstract}
We perform a numerical investigation of the \emph{shaken dynamics}, a parallel Markovian dynamics for spin systems with local interaction and whose transition probabilities depend on two parameters, $q$ and $J$, that tune the geometry of the underlying lattice. We determine a phase transition curve, in the $(q, J)$ plane, separating the disordered phase from the ordered one, study the mixing time of the Markov chain and evaluate the spin-spin correlations as $q$ and $J$ vary. Further, we investigate the relation between the equilibrium measure of the shaken dynamics and the Gibbs measure for the Ising model. Two different approaches are considered for the implementation of the dynamics: a multicore CPU approach, with code written in Julia and a GPU approach with code written in CUDA.

\end{abstract}

\section{Introduction}

The Gibbs sampling of lattice spin models is a major task for statistical mechanics.
The numerical techniques developed for its realization are based mainly on Markov chain dynamics for single and cluster spin flip \cite{Glauber:1963:TDS}\cite{SwendsenWang1987}\cite{PhysRevLett.62.361}, and can be easily implemented by means of random mapping representation \cite{haggstrom2002finite} techniques.

A theory of parallel Markov chains  as a
Probabilistic Cellular Automaton (PCA) dates back to 1989 \cite{lebo89}. These processes are characterized by a factorized transition matrix on the configuration space, and their simulation updating all spins by means of the same random map \cite{shakenDynamicsArxiv2019}. More recently a class of PCAs
where transition probabilities are defined in terms of a \emph{pair Hamiltonian} and where
the spins are simultaneously updated at each time step has been the
subject of several works, e. g.,  \cite{DaiPra2012, Lancia2013} where PCA are exploited to study
the Ising model on planar graphs.
We explore the computational possibilities of this pair Hamiltonian model to generalize the random sampling algorithms for Ising spin systems on a set of two-dimensional lattices.

Formally a PCA is a Markov Chain $({X_n})_{n\in\N}$ whose transition probabilities are such that given two generic configurations $\tau=(\tau_1,\ldots,\tau_k)$ and $\sigma=(\sigma_1,\ldots,\sigma_k)$
\begin{equation}
P\{X_n = \tau | X_{n-1} = \sigma\} =
\prod_{i=1}^k P\{(X_n)_{i} = \tau_{i} | X_{n-1} = \sigma\}
\label{fact}
\end{equation}
so that for each time $n$, the components of the ``configuration'' are independently updated.
From a computational point of view, the evolution of a Markov Chain of this type is well suited to
be simulated on parallel processors and GPUs.

In this framework, a new PCA parameterized by $J$ and $q$, called \emph{shaken dynamics}
has recently been introduced \cite{shakenDynamicsArxiv2019}. The equilibrium measure of the shaken
dynamics has been extensively investigated in \cite{criticalitySquareToHexArxiv2019} and
a critical curve in the plane $(q, J)$ has been explicitly determined.

In particular in \cite{criticalitySquareToHexArxiv2019} a model has been proposed where the configurational variables are split into two groups $\tau=(\tau_1,\ldots,\tau_k)$ and $\sigma=(\sigma_1,\ldots,\sigma_k)$, where $\tau_i,\sigma_i\in\{-1,1\}$ for each $i$, and are arranged on a bipartite graph.
Different interactions among the $\tau$ and $\sigma$ variables give rise to the possibility of interpolation among different lattice geometries.

The PCA we take into account is a parallel and
irreversible version of the heath bath dynamics and is obtained
by concatenating two different update rules \cite{shakenDynamicsArxiv2019}.
By means of the Hamiltonian defined in \cite{criticalitySquareToHexArxiv2019}, which depends on the $(J,q)$ parameters, we identify numerically two regions of the space $(J,q)$ characterized by different behaviors
of the dynamics.

In this framework, a new PCA parameterized by $J$ and $q$,
called \emph{shaken dynamics}, has recently been introduced in \cite{shakenDynamicsArxiv2019}.
The equilibrium measure of the shaken dynamics has been extensively investigated in
\cite{criticalitySquareToHexArxiv2019} and a critical curve in the plane $(q,J)$ has
been explicitly determined.

The elementary step of the shaken dynamics is naturally defined on the a finite subset
$\Lambda$ of the square lattice $\Z^2$ and consists of a sequence of two inhomogeneous
half steps. However, in both \cite{shakenDynamicsArxiv2019,criticalitySquareToHexArxiv2019}
it has been pointed out that the shaken dynamics can be seen as an \emph{alternate dynamics}
on a subset of the honeycomb lattice.
The proposed dynamics, although not faster than {\sl ad hoc} dynamics for specific models, allows to simulate a whole class of statistical mechanics models spanning from the one-dimensional Ising model to the square lattice and hexagonal one across all the intermediate models.

Depending on the values of $J$ and $q$, the shaken dynamics ``formalism'' defined on the square
lattice can be used to simulate a class of Ising models on the honeycomb lattice
(as pointed out in \cite{criticalitySquareToHexArxiv2019}). Some of the values of $J$ and $q$
are particularly interesting because they allow to use the shaken dynamics to simulate
\begin{itemize}
	\item the Ising model on the isotropic hexagonal lattice for $J=q$
	\item (an approximation to) the Ising model on the square lattice for $q >> 1$
	\item the Ising model on a collection of weakly interacting unidimensional systems for small values of $q$.
\end{itemize}

The numerical investigation we put forward is aimed at:
\begin{itemize}
	\item illustrating a simple heuristic method to numerically determine the critical curve
	\item evaluating the mixing time of the chain as a function of $J$ and $q$
	\item studying the spin-spin correlations as a function of $J$ and $q$.
\end{itemize}
Further, for $J = q$ we compare the mixing time of the shaken dynamics with that of a single
spin flip dynamics for the Ising model on the hexagonal lattice and, for $q >> 1$ we also compare
the mixing time of the shaken dynamics with that of a single spin flip and an alternate parallel
dynamics for the Ising model on the square lattice and evaluate the distance between the equilibrium
measure of the shaken dynamics from the Gibbs measure for the Ising model on the square lattice.

%
%


The paper is organized as follows.
In the next section we define the lattice spin model and the PCA dynamics. In Section~\ref{par3}  we present the numerical findings concerning the model under investigation. Finally, in Section~\ref{sec:implementation_details}, we provide some details concerning the two implementations of the dynamics taken into account: the mutlicore CPU one and the GPU one.

\section{The model}
\label{par1}
Consider the Ising Hamiltonian on a graph $G(V,E)$
\begin{equation}
	H_G(\sigma)=-\sum_{(x,y)\in E}J_{xy}\sigma_x\sigma_y
	\label{gs_h1}
\end{equation}
where
 $\sigma_x\in\{-1,1\}$ for all the $x\in V$ and $J_{xy}\in\mathbb{R}^+$.\\\\
We assume that $V=\Lambda_1\cup\Lambda_2$,
where $\Lambda_1$ and $\Lambda_2$ are finite squared subsets of the square lattice
with $L^2$ sites and periodic boundary conditions
\begin{equation}
	\Lambda=\Lambda_1=\Lambda_2=(\mathbb{Z}/L\mathbb{Z})^2
\end{equation}
and all edges in $E$ have one endpoint in $\Lambda_1$ and the other in $\Lambda_2$.
The $\sigma$ and $\tau$ variables denote the Ising configuration on the vertices
of $\Lambda_1$ and $\Lambda_2$.
Each $\sigma_u$, with $u\in\Lambda_1$, can be put in a one-to-one
correspondence with $\tau_u$ with the same index $u\in\Lambda_2$.\\\\
Let $x=(i,j)$ be a vector of coordinate on the torus $({\mathbb Z}/L{\mathbb Z})^2$.
Then
\begin{equation}
	 x^\uparrow=(i,j+1),\quad x^\rightarrow=(i+1,j),\quad x^\downarrow=(i,j-1),\quad x^\leftarrow=(i-1,j)
\end{equation}
are the coordinates of the four points at unit distance from $x$.
Set $J_{xy}=J$ for all $(x,u)\in E$ with $x\neq u$ and $J_{xy}=q$ if $x=y$.

\begin{figure}[tbh]
\centering
\begin{tikzpicture}
   \gridThreeD{0}{0}{black!30};
    \gridThreeD{0}{4.25}{black!30};

    \begin{scope}
        \myGlobalTransformation{0}{0};

        \foreach \x in {3} {
            \foreach \y in {3} {
                \node (thisNode) at (\x,\y){};
				\node (thatNode) at (\x,\y) {};
                {
                    \pgftransformreset
                    \draw[red,myBG]  (thatNode) -- ++(0,4.25);
					\draw[black,myBG]  (thatNode) -- ++(2,4.25);
					\draw[black,myBG]  (thatNode) -- ++(0.8,5.20);
                }
            }
        }
		\node at (-.7,12) {$\tau_x$};
		\node at (-.1,14) {$\tau_{x^\uparrow}$};
		\node at (1.3,12) {$\tau_{x^\rightarrow}$};
		\node at (2.6,3.4) {$\sigma_x$};
		\node at (.5,4) {$\Lambda_1$};
		\node at (-3,13) {$\Lambda_2$};
    \end{scope}

    \begin{scope}
        \myGlobalTransformation{1}{4.25};
    \end{scope}

    \graphThreeDnodes{0}{0};
    \graphThreeDnodes{0}{4.25};

\end{tikzpicture}
\caption{The lattices $\Lambda_1$, $\Lambda_2$ with the $q$ (red) and $J$ (black) interactions.}
\label{gs_doppioReticolo}
\end{figure}
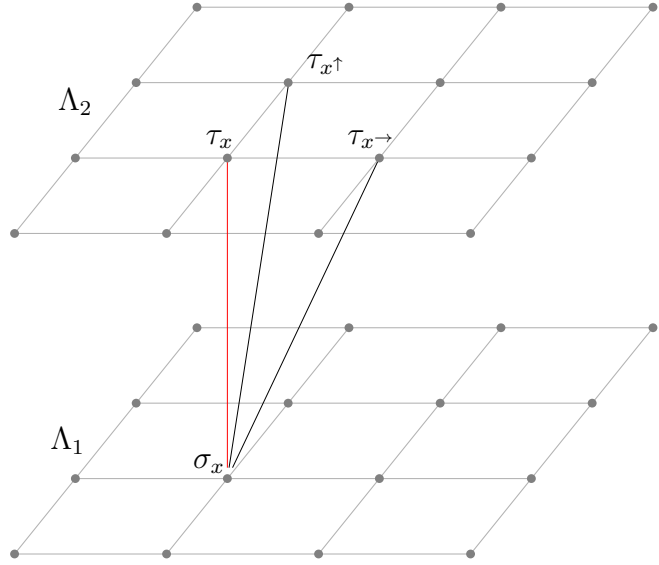

With this notation we obtain the Hamiltonian studied in \cite{criticalitySquareToHexArxiv2019, shakenDynamicsArxiv2019}
\begin{equation}
	\begin{split}
		H(\sigma,\tau)
		&= - \sum_{x \in \Lambda} \left[J \sigma_x(\tau_{x^{\uparrow}} + \tau_{x^{\rightarrow}}) +q\sigma_x \tau_x\right]  \\
		&= - \sum_{x \in \Lambda}\left[J \tau_x(\sigma_{x^{\downarrow}} + \sigma_{x^{\leftarrow}})  +q\tau_x\sigma_x \right]
	\label{gs_ham2}
	\end{split}
\end{equation}
on the pairs of Ising configurations $\sigma$
on $\Lambda_1$ and $\tau$ $\Lambda_2$.
The interactions of this Hamiltonian can be visualized on the induced bipartite graph represented in Fig.~\ref{gs_doppioReticolo} and~\ref{fig:hexSquareLattice}. The parameter $q$ is also referred to as the \emph{self interaction parameter}.

As pointed out in \cite{criticalitySquareToHexArxiv2019} a careful look to the Hamiltonian (\ref{gs_ham2}) and to the graph of Fig.~\ref{gs_doppioReticolo} shows that the bipartite graph is isomorphic to the hexagonal lattice $G^{\varhexagon}(V,E)$ with edges $J$ and $q$ on whose vertices are arranged the variables $\sigma$ and $\tau$ as shown in Fig.~\ref{gs_Honeycombe}.
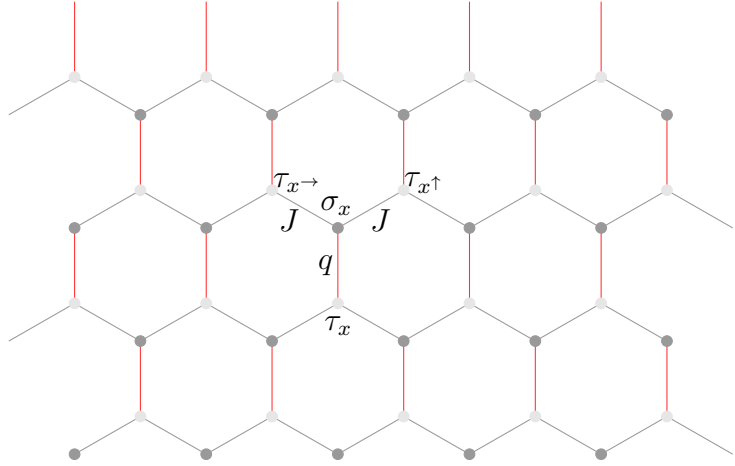
\begin{figure}[tbh]
\centering
\begin{tikzpicture}
  \foreach \i in {0,...,4}
  \foreach \j in {0,...,1} {
  \foreach \a in {-30,210} \draw[gray!80] (2*sin{60}*\i,3*\j) -- +(\a:1);
  \foreach \a in {-30,210} \draw[gray!80] (2*sin{60}*\i+cos{30},3*\j-1-sin{30}) -- +(\a:1);
  \foreach \a in {90} \draw[red!80] (2*sin{60}*\i,3*\j) -- +(\a:1);
  \foreach \a in {90} \draw[red!80] (2*sin{60}*\i+cos{30},3*\j-1-sin{30}) -- +(\a:1);
}
  \foreach \i in {0,...,4}
  \foreach \j in {0,...,1} {
  \filldraw [gray!20] (2*sin{60}*\i,3*\j) circle (2pt);
  \filldraw [gray!80] (2*sin{60}*\i,3*\j-2) circle (2pt);
  \filldraw [gray!20] (2*sin{60}*\i+cos{30},3*\j-1-sin{30}) circle (2pt);
  \filldraw [gray!80] (2*sin{60}*\i+cos{30},3*\j-sin{30}) circle (2pt);
}
\draw[] (3.8,1.3) node[left,black]{$\sigma_x$};
\draw[] (3.8,-0.3) node[left,black]{$\tau_x$};
\draw[] (3.4,1.6) node[left,black]{$\tau_{x^{\rightarrow}}$};
\draw[] (5.0,1.6) node[left,black]{$\tau_{x^{\uparrow}}$};
\draw[] (3.55,0.5) node[left,black]{$q$};
\draw[] (3.1,1.1) node[left,black]{$J$};
\draw[] (4.3,1.1) node[left,black]{$J$};
\end{tikzpicture}
\caption{The hexagonal graph $G^{\varhexagon}(\Lambda_1\cup\Lambda_2,\{J,q\})$}
\label{gs_Honeycombe}
\end{figure}
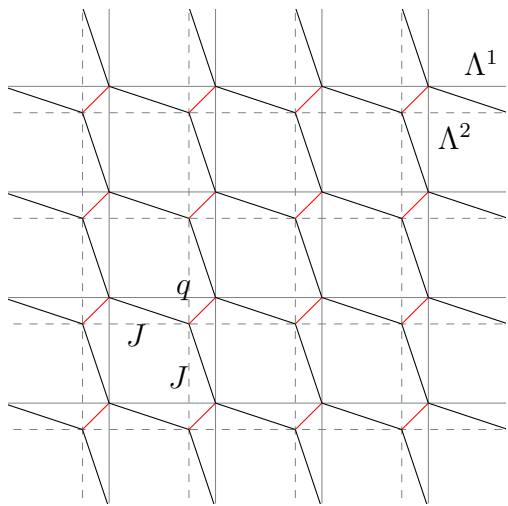
\begin{figure}[tbh]
\centering

\begin{tikzpicture}[scale=0.7]

	\tikzmath{
		\xs=0.5;
		\ys=0.5;
	}
	\clip (0.1, 0.1) rectangle (9.45, 9.45);

	\draw[step=2cm,gray, help lines] (0.1,0.1) grid (9.9,9.9);
	\draw[step=2cm,gray, help lines, dashed, shift={(-\xs, -\ys)}] (0.1,0.1) grid (9.9,9.9);

	\foreach \x in {0, 2, 4, 6, 8}
		\foreach \y in {0, 2, 4, 6, 8}
			{\draw[red, thin] (\x,\y) -- (\x-\xs, \y-\ys);
			 \draw[black, thin] (\x,\y) -- (\x-\xs, \y+2-\ys);
			 \draw[black, thin] (\x,\y) -- (\x+2-\xs, \y-\ys);
			}

		\path(4,4) -- node [anchor=south east]{$q$} (4-\xs, 4-\xs);
		\path(2,4) -- node [anchor=north east, pos=0.6]{$J$} (4-\xs, 4-\ys);
		\path(4,2) -- node [anchor=north east, pos=0.6]{$J$} (4-\xs, 4-\ys);

		\path(8,8) -- node [color=black, anchor=south]{$\Lambda^1$} (10,8);
		\path[xshift=-\xs cm, yshift=-\ys cm](8,8) -- node [color=black, anchor=north]{$\Lambda^2$} (10,8);

\end{tikzpicture}
\caption{A representation of the hexagonal graph $G^{\varhexagon}(\Lambda_1\cup\Lambda_2,\{J,q\})$ that highlights the
relation with the two square lattices $\Lambda_{1}$ and $\Lambda_{2}$}
\label{fig:hexSquareLattice}
\end{figure}
The Gibbs measure at temperature $1/\beta$ for the Hamiltonian \eqref{gs_ham2} is
\begin{equation}
	\pi_2(\sigma,\tau)={e^{-\beta H(\sigma,\tau)}
	\over\sum_{(\sigma,\tau)\in{\mathcal X}\times{\mathcal X}}e^{-\beta H(\sigma,\tau)}}
\end{equation}
where ${\mathcal X}\times{\mathcal X}=\{-1,1\}^{|\Lambda|}\times\{-1,1\}^{|\Lambda|}$ is the configuration space of the variable $(\sigma,\tau)$.
The critical value of $\beta_c$ separates the ordered phase where all the spin have the same probability to take the values $+1$ or $-1$ from the ordered phase where the measure is polarized \cite{Gallavotti1972}.

\begin{figure}[tb]
\centering

\begin{tikzpicture}
\begin{axis}[
x=60,
y=60,
xlabel={$q$},
ylabel={$J_c(q)$},
samples=41,
grid,
domain=-4:4,
legend pos=outer north east,
samples=1500,
xmin=0,
xmax=3.5,
ymin=0,
ymax=3.5
]
\draw[] (0.8,0.7) node[right,black]{$0.6585$};
\draw[] (3.3,0.6) node[left,black]{$0.4407$};
\addplot [domain=0:4,mark=none,black] {arctanh(sqrt(tanh(\x)*tanh(\x)+1)-tanh(\x))};
\addplot [domain=0:4,mark=none,red] {x};
\addplot [domain=0:4,mark=none, red, dashed] {0.4406867};
\end{axis}
\end{tikzpicture}
\caption{The critical curve $J_c(q)$}
\label{Jqplot}
\end{figure}
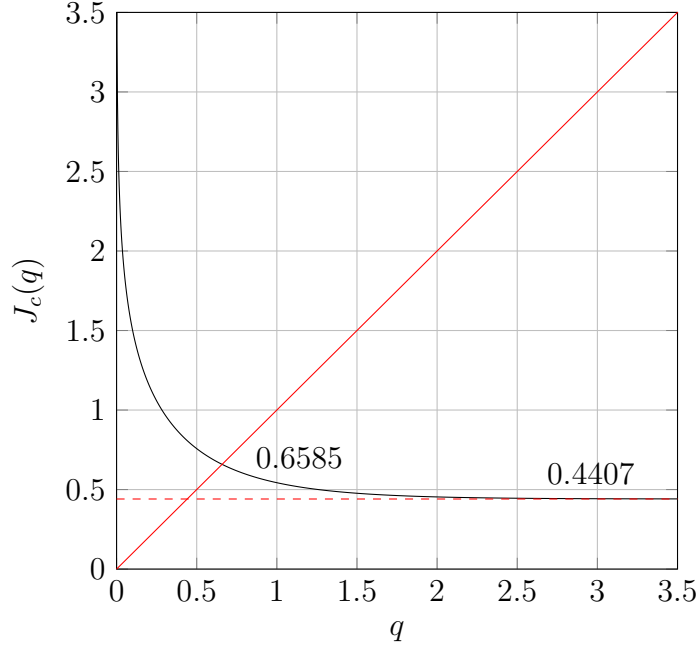

Rescaling the interactions $J$ and $q$ by $\beta$
\begin{equation}
	\beta J\rightarrow J,\qquad \beta q\rightarrow q
\end{equation}
it has been proven in  \cite{criticalitySquareToHexArxiv2019} that there exists a function $J_c(q)$, shown in Fig.~\ref{Jqplot}, which separate the ordered phase from the disordered one.

The partition function of the Ising model on the honeycomb lattice $G^{\varhexagon}$ is
\begin{equation}
	Z(J,q)
	=\sum_{(\sigma,\tau)\in{\mathcal X}\times{\mathcal X}}
		\prod_{u\in\Lambda}\cosh^2 J\cosh q\big(1+\sigma_x\tau_{x^\uparrow}\tanh J\big)
			\big(1+\sigma_x\tau_{x^\rightarrow}\tanh J\big)\big(1+\sigma_x\tau_x\tanh q\big)
\end{equation}\ \\
The graph $G^{\varhexagon}$ is a weighted planar graph, non degenerate, finite and doubly periodic.
The periodic boundary conditions for $\Lambda_1$ and $\Lambda_2$ guarantee that the graph $G^{\varhexagon}$ is  immersed in the torus.

Introducing the following notation on the hexagonal lattice
\begin{equation}
	J_e\equiv\left\{
	\begin{split}
	 & J \text{ if } e=(x,x^\uparrow) \text{ or } e=(x,x^\rightarrow)\\
	 & q \text{ otherwise }
	\end{split}
	\right.
\end{equation}
the critical curve $J_c(q)$ for the Hamiltonian (\ref{gs_ham2}) is the unique solution for $J,q>0$ of the equation
\begin{equation}
	\sum_{\gamma\in{\mathcal E}_0(G)}\prod_{e\in\gamma}\tanh J_e=\sum_{\gamma\in{\mathcal E}_1(G)}\prod_{e\in\gamma}\tanh J_e
\label{gs_eq_crit}
\end{equation}
where ${\mathcal E}_0(G)$ is the set of even subgraphs of $G^{\varhexagon}$ winding an even number of times around each direction of the torus, and ${\mathcal E}_1(G)={\mathcal E}(G)\setminus{\mathcal E}_0(G)$ \cite{criticalitySquareToHexArxiv2019}\cite{CimasoniDC2013}.\\
The explicit form of the equation (\ref{gs_eq_crit}) is
\begin{equation}
1=	2\tanh J\tanh q+\tanh^2 J
\label{eqcrit}
\end{equation}
The solution of eq.(\ref{eqcrit}) with respect to the $J$
\begin{equation}
\label{J_di_q}
	J_c(q)=\tanh^{-1}\big(\sqrt{\tanh^2q+1}-\tanh q\big)
\end{equation}
is plotted in Fig.~\ref{Jqplot}.

We observe that
\begin{equation}
	\lim_{q\rightarrow\infty}J_c(q)=\tanh^{-1}(\sqrt{2}-1)=0.4406867
\end{equation}
is the critical value of $\beta$ for the Ising model on the lattice square, while on the point $J_c(q)=q$ the equation (\ref{J_di_q}) gives the critical value for the Ising model on the hexagonal lattice $J=q=0.6585$.
If $q\rightarrow0,J\rightarrow\infty$ there are no phase transitions as in the unidimensional Ising model.

Following \cite{shakenDynamicsArxiv2019} we let the system evolve as a
Markov chain where the spins in $\Lambda_1$ and
in $\Lambda_2$ are alternatively updated
with a probability proportional to the exponential of the
Hamiltonian of the target configuration in ${\mathcal X}\times{\mathcal X}$.

More precisely, using the notation
\begin{equation}
	\begin{split}
\rightField{x}{\sigma} &= J(\spinUp{x}{\sigma} + \spinRight{x}{\sigma}) + q \spin{x}{\sigma}\\
\leftField{x}{\sigma} &= J(\spinDown{x}{\sigma} + \spinLeft{x}{\sigma}) + q \spin{x}{\sigma}
	\end{split}
	\label{eq:urdl}
\end{equation}
we consider a Markov chain on ${\mathcal X}\times{\mathcal X}$
with transition probabilities given by
\begin{align}\label{eq:leftTrans}
    P((\sigma, \tau),(\sigma, \tau\prm))
    = \frac{e^{-H(\sigma,\tau\prm)}}{Z_{\sigma}}
    = \frac{e^{-\sum_{u\in \Lambda}\leftField{u}{\sigma} \spin{u}{\tau\prm}}}{Z_{\sigma}}
    = \prod_{u\in\Lambda}\frac{e^{\leftField{u}{\sigma} \spin{u}{\tau\prm}}}{2 \cosh{\leftField{u}{\sigma}}}
\end{align}
at odd times and
\begin{align}\label{eq:rightTrans}
    P((\sigma, \tau),(\sigma\prm, \tau))
    = \frac{e^{-H(\sigma\prm,\tau)}}{Z_{\tau}}
    = \frac{e^{-\sum_{u\in \Lambda}\rightField{u}{\tau} \spin{u}{\sigma\prm}}}{Z_{\tau}}
    = \prod_{u\in\Lambda}\frac{e^{\rightField{u}{\tau} \spin{u}{\sigma\prm}}}{2 \cosh{\rightField{u}{\tau}}}
\end{align}
at even times where
$Z_\sigma=\sum_{\eta\in{\mathcal X}} e^{-H(\sigma,\eta)}$
and
$Z_\tau=\sum_{\eta\in{\mathcal X}} e^{-H(\eta,\tau)}$.

The factorization in  eq.~\eqref{eq:leftTrans}
and~\eqref{eq:rightTrans}
and the mutual dependence of the variables $\sigma$ and $\tau$
makes it quite easy the parallel numerical implementation
of this dynamics.
In particular, to simulate the evolution of the chain
it is possible to sample the value
$\zeta \in \{-1, 1\}$ of the
spin at site $u$ with probability
$P(\tau\prm_{u} = \zeta|\sigma) =
        \frac{e^{\zeta \leftField{u}{\sigma} }}{2\cosh{\leftField{u}{\sigma}}}$
at odd times and
$P(\sigma\prm_{u} = \zeta|\tau) =
        \frac{e^{\zeta \rightField{u}{\tau} }}{2\cosh{\rightField{u}{\tau}}}$
at even times independently for all $u \in \Lambda$.

In this framework, the shaken dynamics introduced in \cite{shakenDynamicsArxiv2019}
is obtained by looking at the evolution of the spin configuration in $\Lambda_1$.
In other words, the shaken dynamics is the marginal of the alternate dynamics defined
by eq.~\eqref{eq:leftTrans} and~\eqref{eq:rightTrans} and the shaken transition probabilties are
\begin{align*}
    P^{\square}(\sigma, \sigma\prm) =
    \sum_{\tau} \frac{e^{-H(\sigma,\tau)}}{Z_{\tau}} \frac{e^{-H(\sigma\prm,\tau)}}{Z_{\tau}}
\end{align*}

In \cite{shakenDynamicsArxiv2019} it has been proven that the equilibrium measure of this dynamics is\
\begin{align*}
\pi_{s}(\sigma)=\frac{{Z}_\sigma}{Z}
\end{align*}
and $Z = \sum_{\sigma} Z_{\sigma}$.

In the remainder of the paper, we use the wording \emph{shaken dynamics} when we are interested in the evolution on the sub--lattice $\Lambda_1$ whereas we call the dynamics on the hexagonal lattice subject to the transition probabilities \eqref{eq:leftTrans} and~\eqref{eq:rightTrans} the \emph{alternate parallel dynamics} (on the hexagonal lattice).


\section{Simulation results}\label{par3}
\subsection{Numerical estimation of critical curve}
As stated before, the critical curve \eqref{J_di_q} is the function that separates the ordered and the disordered phases.
Above this line the values of the spins tend to be highly correlated
whereas on the opposite side the value assumed by each spin is weakly dependent on the
values taken by other spins.
To determine whether the system is in the ordered or disordered phase we compute, over a large number of iterations, the average and the variance of the magnetization on one of the two
layer $\Lambda_i$ is where the magnetization $m$ is defined as
\begin{align}
	m = \frac{1}{|\Lambda_{i}|} \sum_{x \in \Lambda_i}\sigma_x
\end{align}
By Theorem~2.1 in \cite{criticalitySquareToHexArxiv2019} $\pi_{s}(m) = \pi_{2}(m)$, that is, the average magnetization
(in $\Lambda_{1}$) of the shaken dynamics is the same as the average magnetization of the parallel alternate dynamics (on the
hexagonal lattice $\Lambda_{1} \cup \Lambda_{2}$).

We take $\Lambda$ to be a $200 \times 200$ torus and simulate the
evolution of the shaken dynamics starting from configuration $\sigma_0 = \{-1, -1, \ldots -1\}$ for
$(J, q) \in \{(0, 2) \times (0,2)\}$ on a $80 \times 80$ grid. We first let the system run for a warm-up time of 300000 steps
and then record the average and the variance of the magnetization
for 300000 additional steps.

\begin{figure}[h!]
   \centering
   \subfigure[]{
      \includegraphics[width=0.45\linewidth]{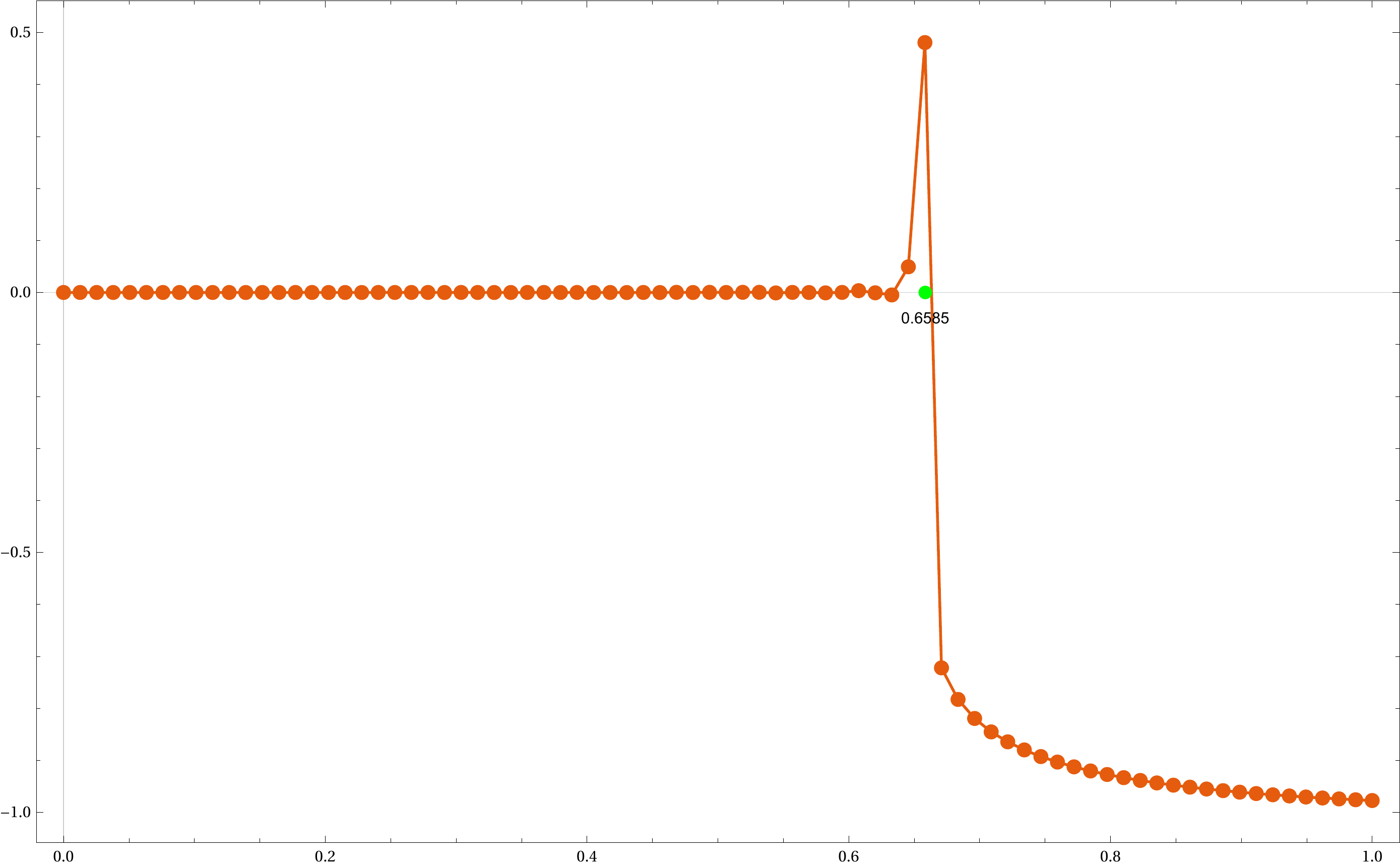}
   }
   \hspace{\fill}
   \subfigure[]{
      \includegraphics[width=0.45\linewidth]{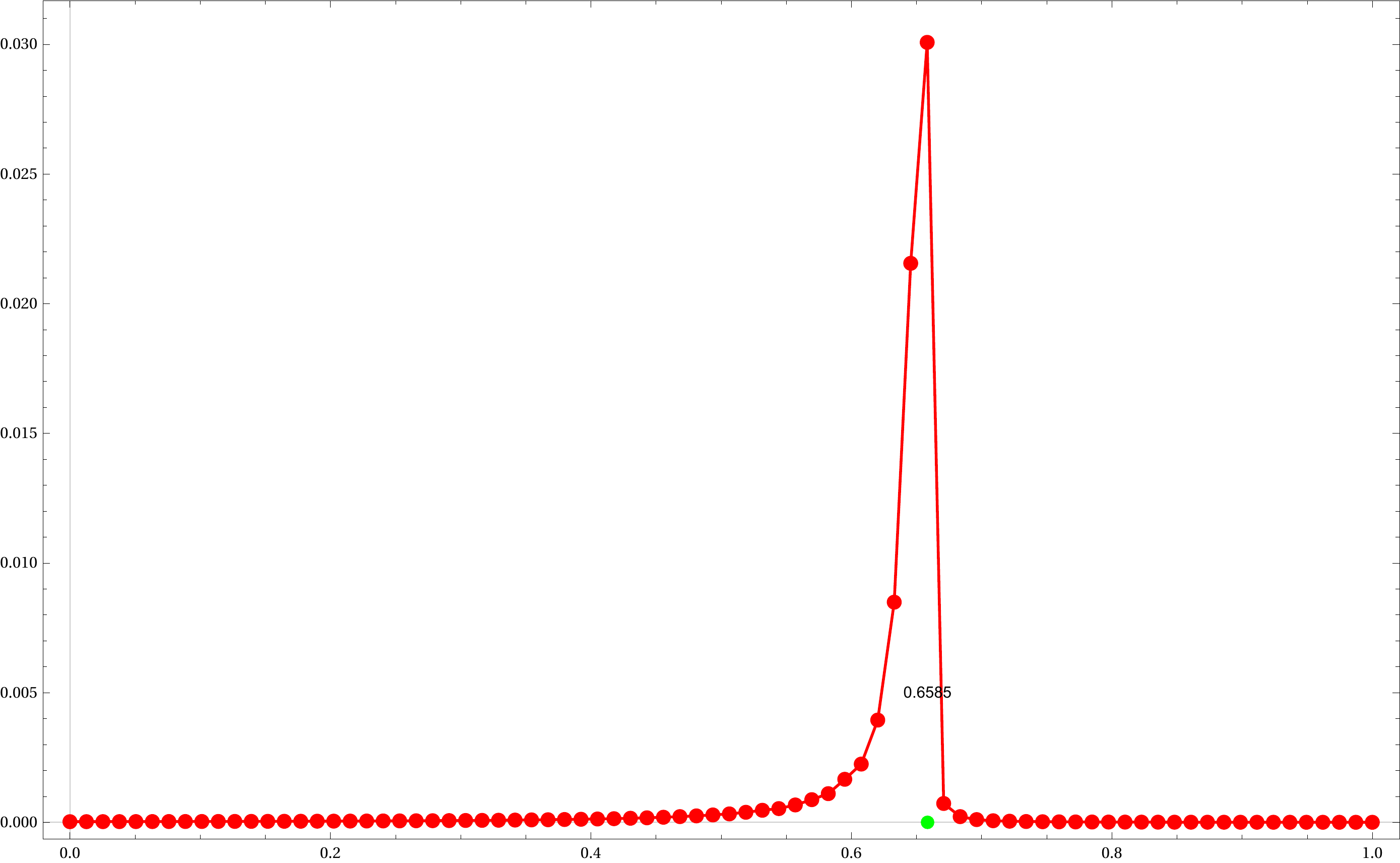}
   }
   \caption{Average (a) and variance (b) of the magnetization as a function of $J$ for $q=0.6585$}
   \label{fig:media_varianza}
\end{figure}

Figure~\ref{fig:media_varianza} shows the average and the variance of
the magnetization as a function of $J$ for $q = 0.6585$.
It is evident that the average magnetization has a sharp transition around the point
$J = 0.6585$ which is the critical value of $J$ for the Ising model on the honeycomb lattice.
Around the same point the variance of the magnetization has a spike whereas
it is negligible for values of $J$ far from the critical point.

\begin{figure}[h!]
   \centering
   \subfigure[]{
      \includegraphics[width=0.46\textwidth]{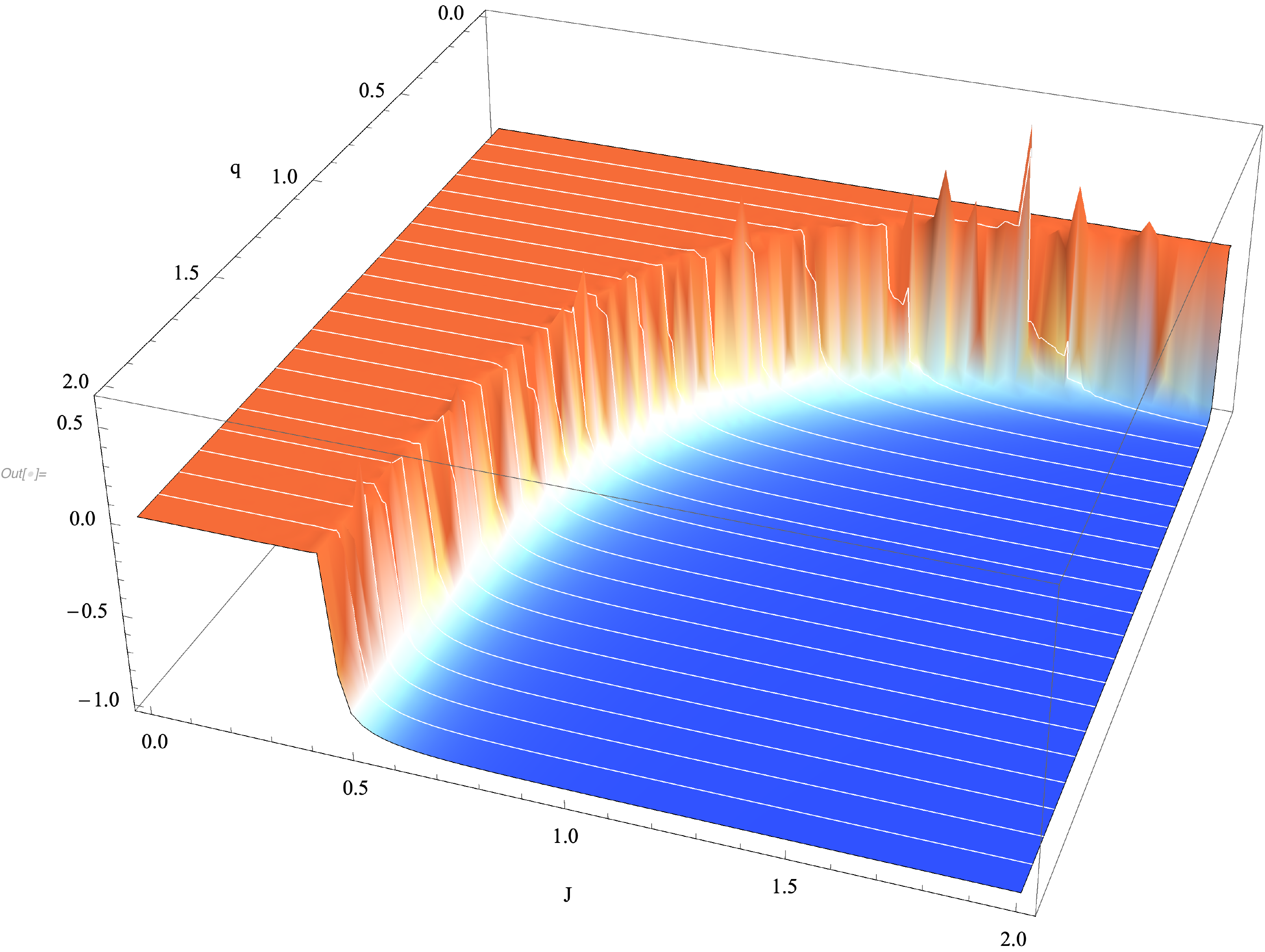}
   }
   \hspace{\fill}
   \subfigure[]{
      \includegraphics[width=0.46\textwidth]{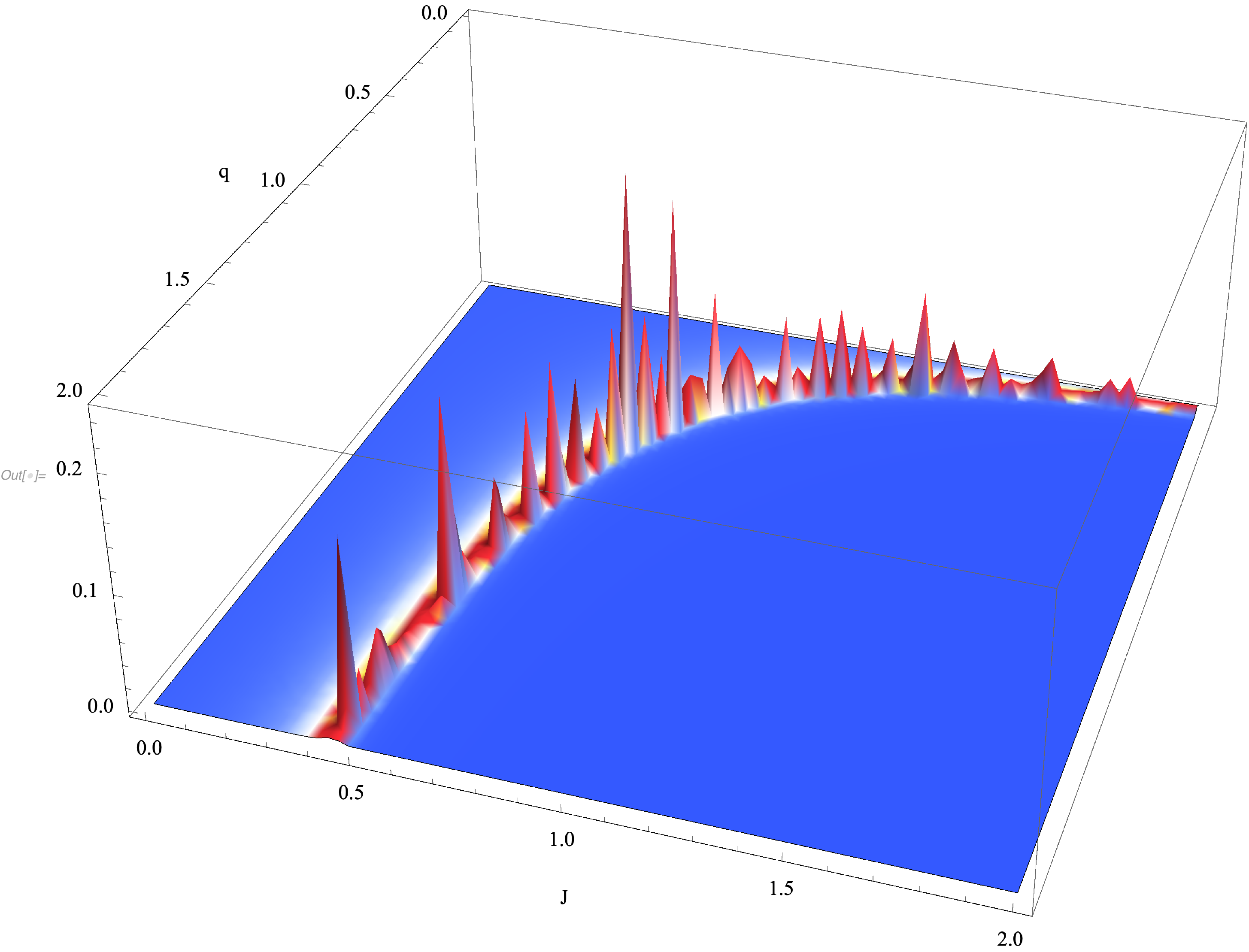}
   }
   \caption{Average (a) and variance (b) of the magnetization on the whole $(q,J)$ grid}
   \label{fig:niagara}
\end{figure}

The results obtained on the whole grid $(q,J)$ are summarized in
Figure~\ref{fig:niagara}.

\begin{figure}[h!]
   \centering
   \includegraphics[width=0.8\linewidth]{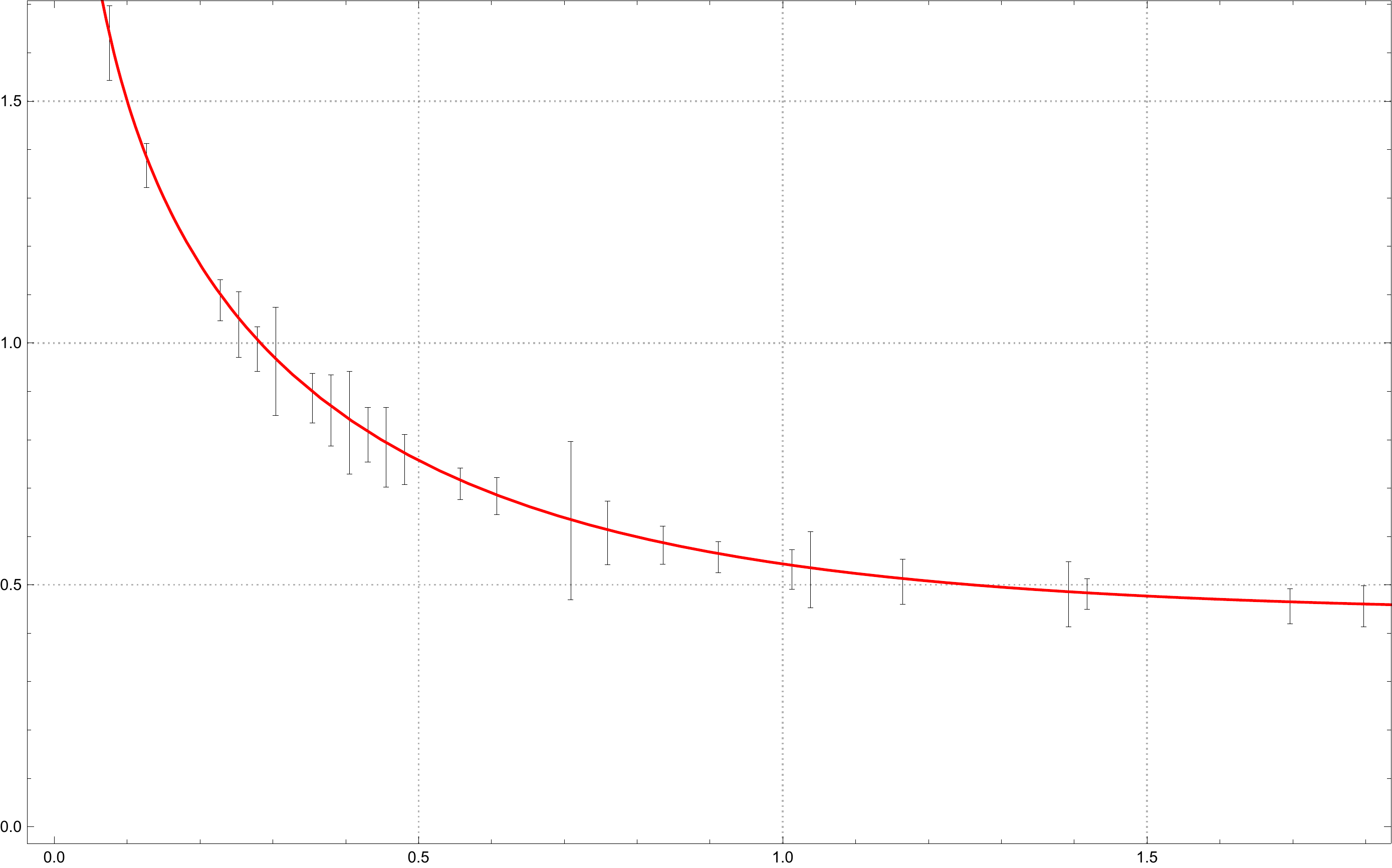}
   \caption{The bars are centered at those points in the $(q,J)$ plane for which the variance
   of the magnetization is sufficiently large ($\ge 0.03$). The length of the bars is proportional
   to the variance of the magnetization.}
   \label{fig:critica}
\end{figure}

It is known that, at equilibrium, the average value of the magnetization fluctuates heavily only
close to the critical value of the interactions (see \cite{ruelleBook} for a reference).
Figure~\ref{fig:critica} shows that the variance of the magnetization is significantly different from zero only for points of the $(q,J)$ plane in the vicinity of the pins on the curve \eqref{J_di_q}.
This show that, even for a small lattice, the magnetization fluctuates only close to the critical
line and for the whole class of Ising models that can be described tuning the values of $J$ and $q$.

\subsection{Coalescence times and perfect sampling}
To assess whether the number of steps for which a Markov chain is run is large enough for its distribution to be close to the equilibrium distribution, it is convenient to look at its \emph{mixing time}. For a Markov chain $({X_n})_{n\in\N}$ with state space $\cX$ and stationary distribution $\pi$, the mixing time is defined as
\begin{align*}
	\tmix = \tmix(\epsi) = \min\{n > 0 : \tvar{\mu_\sigma^n}{\pi} < \epsi \, \forall \, \sigma \in \cX\}
\end{align*}
where $\mu_\sigma^n$ is the distribution of $X_n$ conditioned on $X_0 = \sigma$, $\tvar{\mu}{\nu}$ denotes the total variation distance between the probability measures $\mu$ and $\nu$ and $\epsi$ is some ``small'' number (for instance $e^{-1}$). For a reference on mixing times see, for instance, \cite{levinPeres}.
Determining useful bounds for the mixing time of a Markov chain is, in general, a quite challenging task. However, indication on the mixing time of a Markov chain can be gathered looking at the \emph{coalescence times} (see \cite{haggstrom2002finite} for a reference).

Consider two Markov chains $({X_n})_{n\in\N}$ and $({Y_n})_{n\in\N}$ living on the same state space $\cX$ and consider the coupling $({Z_n})_{n\in\N} = (X_n, Y_n)$ obtained by letting $X_n$ and $Y_n$ to evolve with the same update function and the same sequence of random numbers (for an introduction on the coupling method see \cite{lindvall2012lectures}). Further assume that the update function is chosen is a way such that $P_Z(X_n = Y_n) \to 1$ as $n \to \infty$.

We define the coalescence time $T$ between $X_n$ and $Y_n$ as $T = \min\{n\in\N : X_n = Y_n\}$. Note that, since $X_n$ and $Y_n$ evolve with the same update function and the same sequence of random numbers $X_n = Y_n$ for all $n > T$. This definition extends naturally to a collection of $K$ chains $X_n^k$ with $k \in 1\ldots K$.

The mixing time of the chain $({X_n})_{n\in\N}$ is estimated by the coalescence time of the chains
$({X_n^k})_{n\in\N}$ for $k = 1\ldots |\cX|$, all defined on the state space $\cX$, where chain
$X_n^k$ has initial distribution concentrated on state $k$.

To effectively determine the coalescence time of the shaken dynamics, however, it is not necessary to run $2^{|\Lambda|}$ copies of the Markov chains, but it is possible to use the so called sandwiching technique since the shaken dynamics preserves the \emph{partial ordering} between configurations\footnote{$\sigma \ge \eta$ if, for all $u \in \Lambda$, $\{\eta_{u} = +1\} \Rightarrow \{\sigma_{u} = +1$\}} . In other words, it can be directly checked that if $X_0^k \le X_0^l$ than $X_n^k \le X_n^l$ for all $n > 0$ (see, again, \cite{haggstrom2002finite}, for a reference). To determine the coalescence time it is therefore sufficient to look at the coalescence times of two chains starting, respectively, from $\sigma^{\text{top}} = \{1, 1, \ldots, 1\}$ (the largest possible configuration) and $\sigma^{\text{bot}} = \{-1, -1, \ldots, -1\}$ (the smallest possible one).

Further note that, leveraging on coupling between Markov chains it is possible to perform an unbiased sampling from the equilibrium distribution of a Markov chain using the Propp--Wilson algorithm, introduced in \cite{Propp-wilson}, which requires two copies of the Markov chain to be run with the same update function and the same sequence of random numbers.

We studied the coalescence times of the shaken dynamics.
The simulations were run taking $\Lambda$ to be a $32 \times 32$ square lattice.
This means that the induced hexagonal lattice $\Lambda_1 \cup \Lambda_2$ has $32 \times 32 \times 2$ points.

\begin{figure}[h!]
   \centering
      \includegraphics[width=0.7\linewidth]{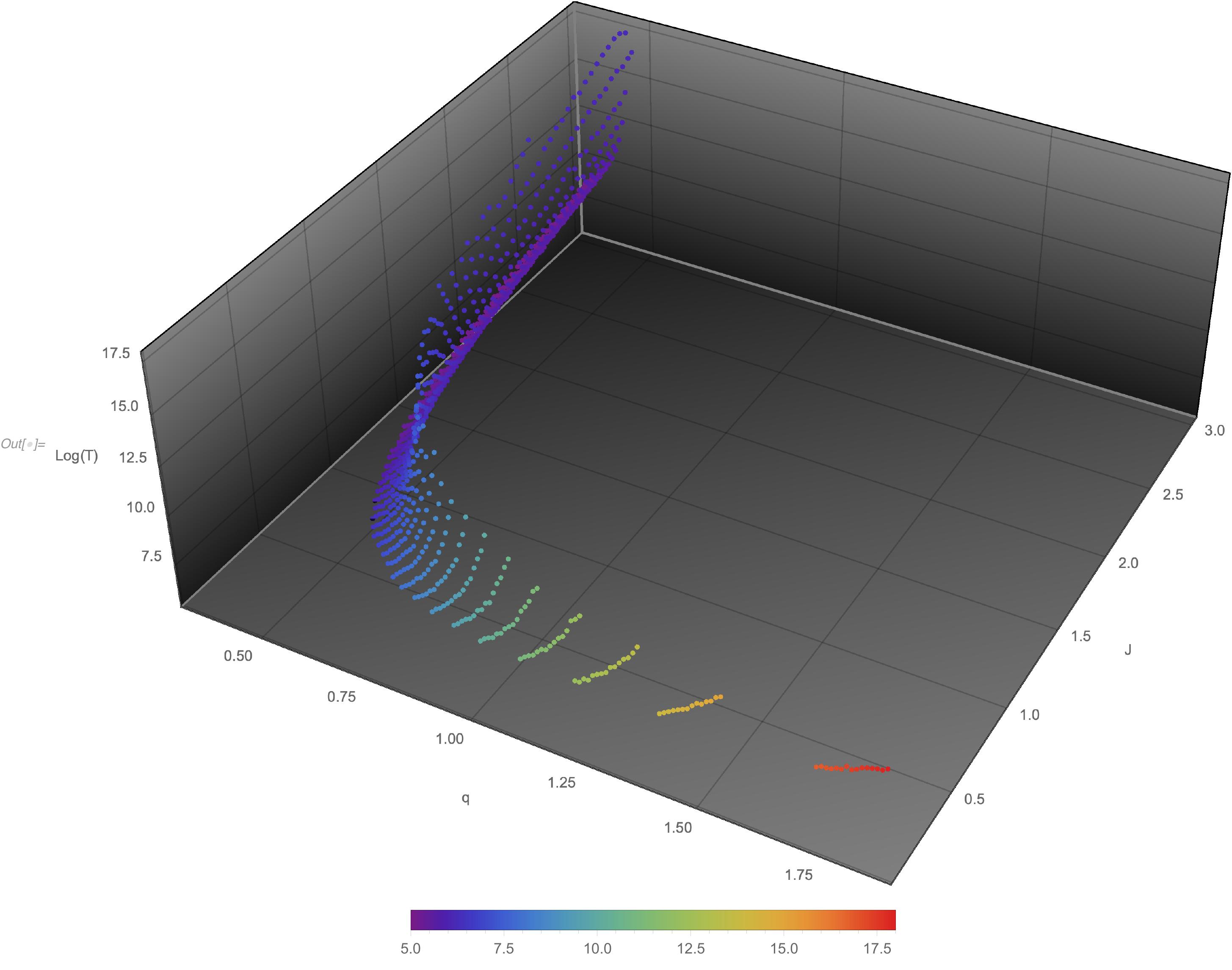}
   \caption{Logarithm of the average coalescence time for values of $J$ and $q$ close to the critical curve}
   \label{fig:scatter_coalescence}
\end{figure}

We computed the average coalescence time for values of $J$ and $q$ close to the critical line $J_c(q)$. The results obtained are summarized in Fig.~\ref{fig:scatter_coalescence}.

For $J=q$, the shaken dynamics is the marginal of the alternate dynamics
on the isotropic hexagonal lattice.
More properly, pairs of configurations $(\sigma, \tau)$ with $\tau$ the configuration obtained
from $\sigma$ by performing the first half step of the shaken dynamics can be
regarded as spin configurations on the honeycomb lattice.
The equilibrium distribution of these pairs is the Gibbs measure of the Ising model on the isotropic
hexagonal lattice (see Theorem~2.1 in \cite{shakenDynamicsArxiv2019}). Therefore it makes sense to compare the mixing time of the shaken dynamics
with the mixing time of a single spin flip dynamics defined on the hexagonal lattice and whose
stationary distribution is the Gibbs measure.
As a reference we take the heat bath dynamics defined as follows:
\begin{align*}
   P(\sigma, \sigma\prm) = \begin{cases}
		\frac{1}{|\Lambda|} \frac{e^{h_{x}(\sigma)\sigma_i}}{2 \cosh(e^{h_{x}(\sigma)})}& \text{if } \sigma\prm = \sigma^{x} \\
		1 -\sum_{x\in \Lambda} P(\sigma, \sigma\prm) & \text{if } \sigma = \sigma\prm \\
		0 & \text{otherwise}
	\end{cases}
\end{align*}
where $\sigma^{x}$ is the configuration obtained from $\sigma$ by flipping the spin
at site $u$ and $h_{u}(\sigma) = \sum_{y\sim x} J \sigma_y$.
Also the heat bath dynamics preserves the partial ordering between configurations and, hence, also in this case it is sufficient to simulate the evolution of two chains one starting from all spins set to $+1$ and one starting from all spins set to $-1$.

Note that it is possible to argue that the parallel alternate dynamics studied here is a parallel version of the single spin flip heat bath described above.

\begin{figure}[tbh]
    \centering
    \includegraphics[width=\linewidth]{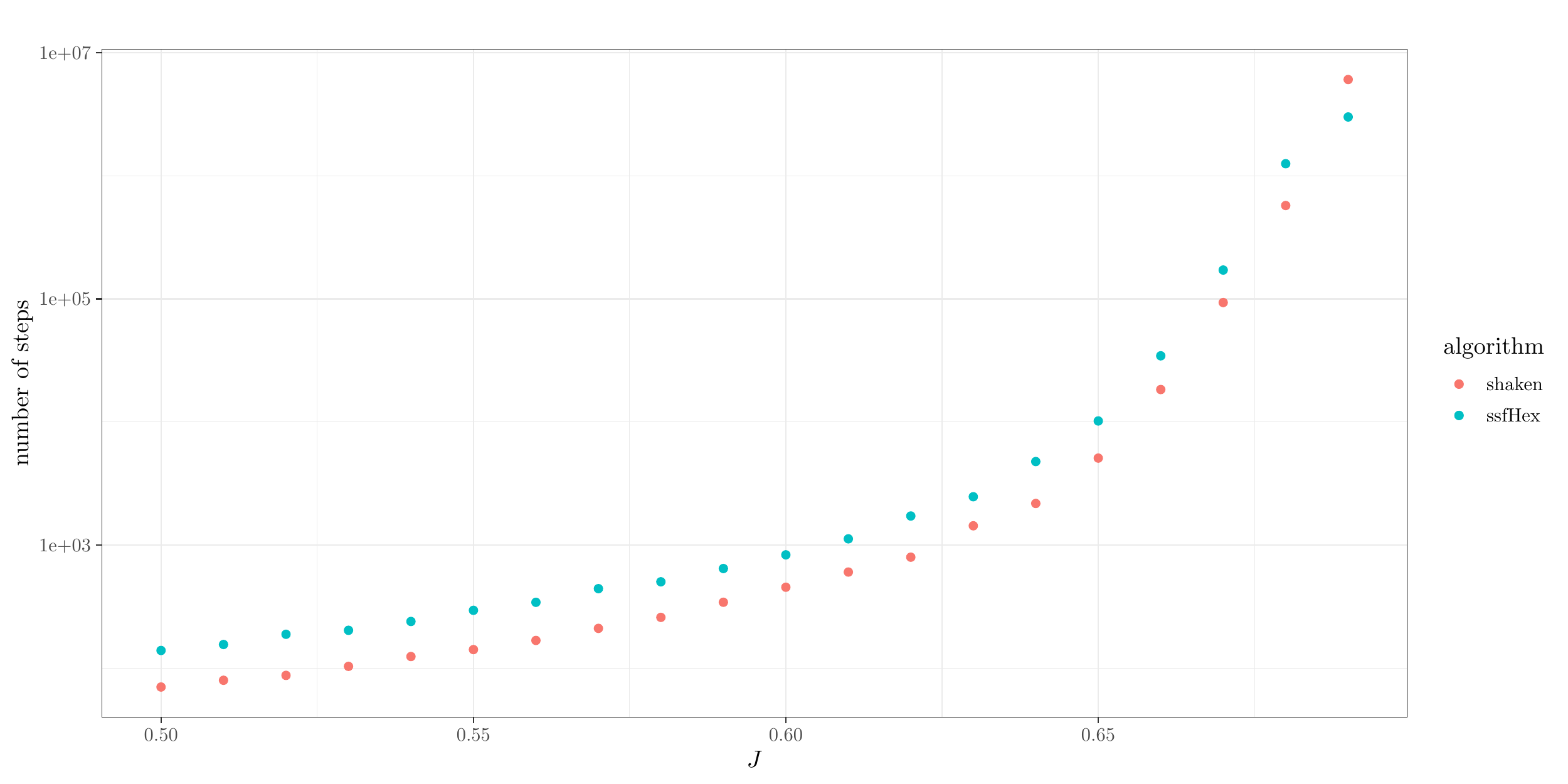}
    \caption{Sample average of the coalescence time for $J=q$ (hexagonal lattice)}
    \label{fig:hex_coaltime}
\end{figure}

The results obtained, for several values of $J$ (and, consequently, $q$) are presented in Fig.~\ref{fig:hex_coaltime}. Note that for the single spin flip dynamics the value shown in the chart is the number of steps divided by $2|\Lambda|$ so that, for both algorithms, we are comparing the total number of ``attempted spin flips''.

It appears that the parallel alternate dynamics is faster than the single spin flip one even if the single spin flip one is ``renormalized'' with the volume of the box as described above.

\begin{figure}[tbh]
    \centering
    \includegraphics[width=1.\textwidth]{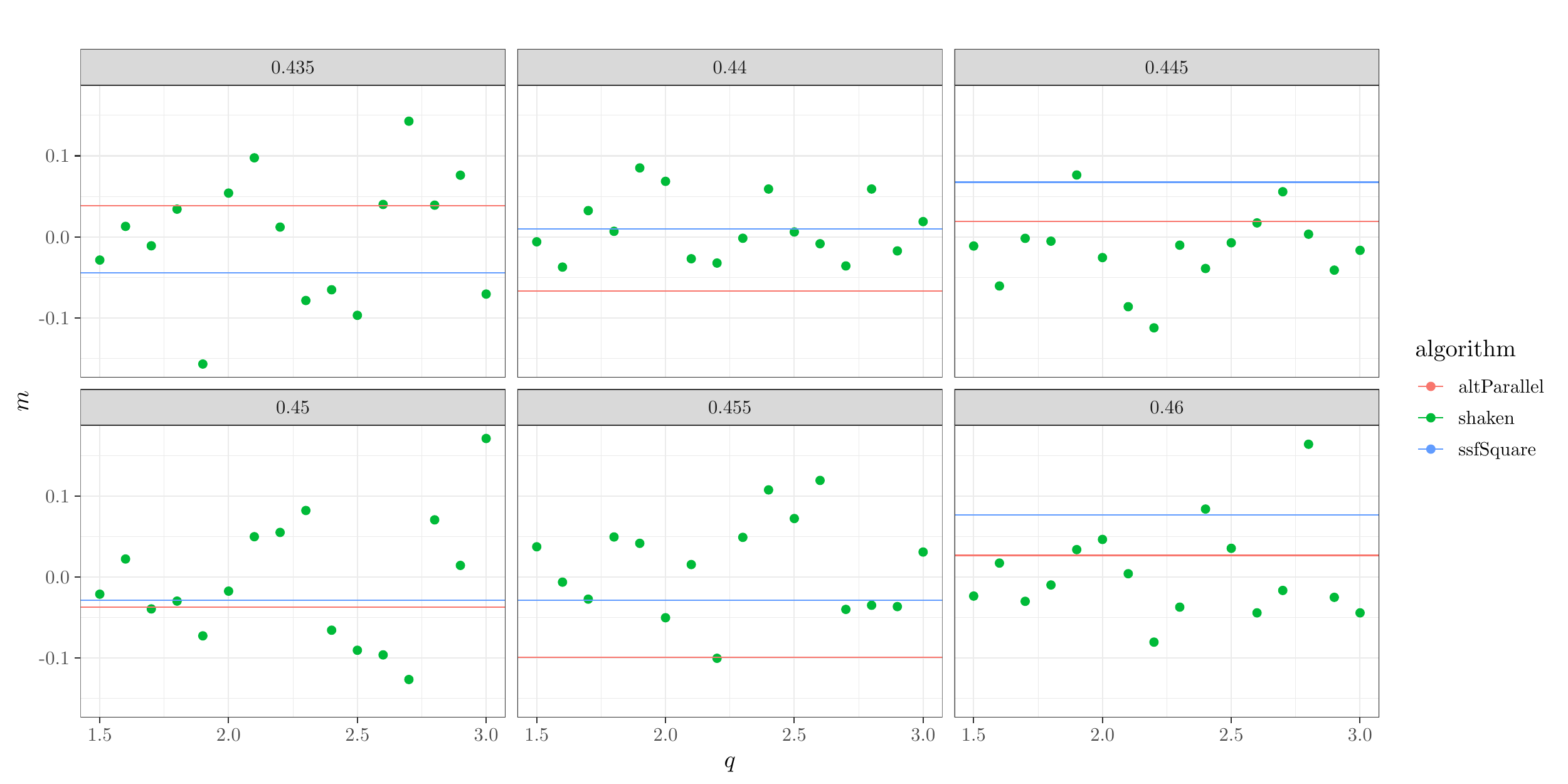}
    \caption{Sample average of the magnetization for several values of $J$}
    \label{fig:avg_mag_square}
\end{figure}

\begin{figure}[tbh]
    \centering
    \includegraphics[width=1.\textwidth]{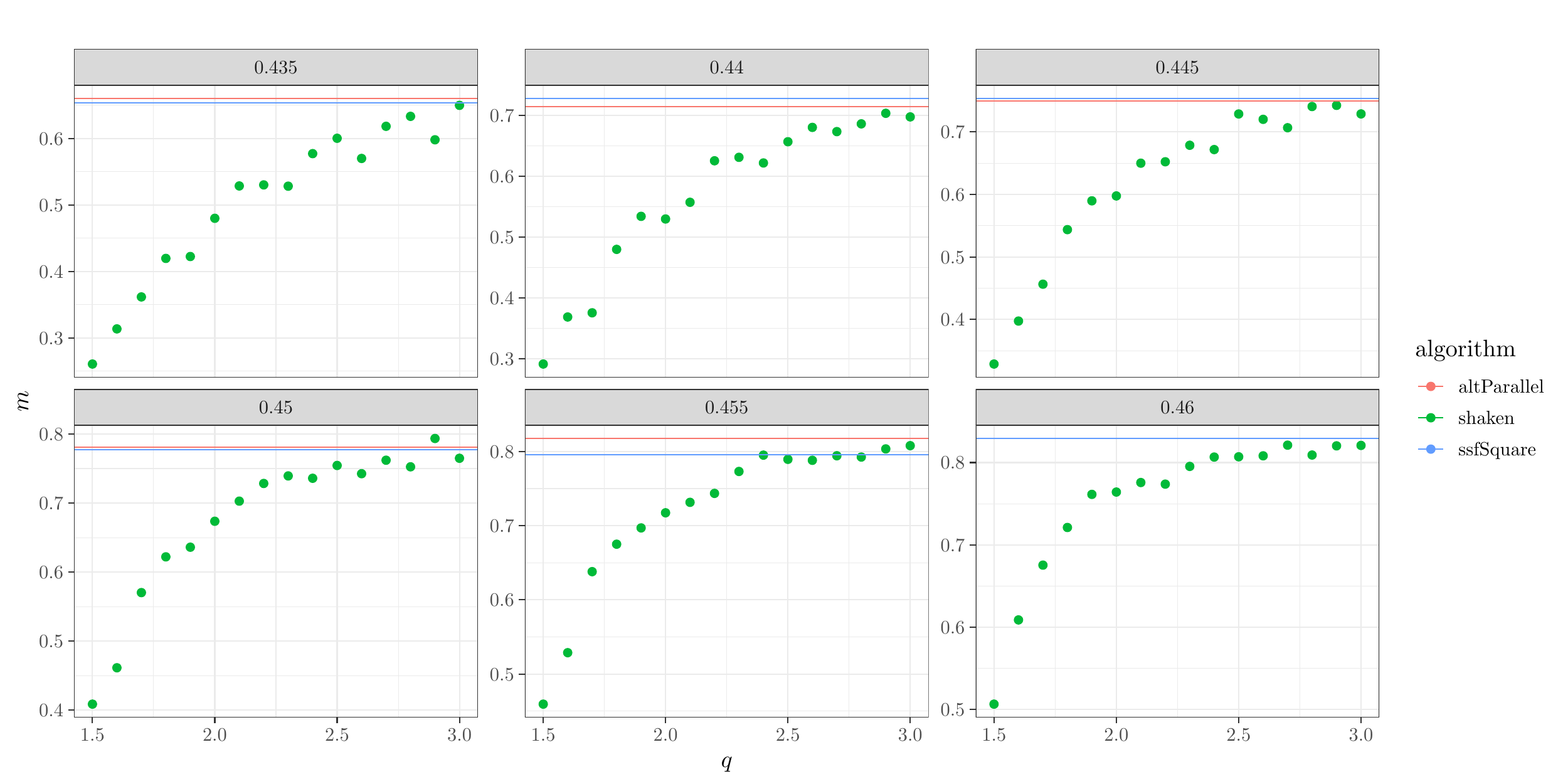}
    \caption{Sample standard deviation of the magnetization for several values of $J$}
    \label{fig:sd_mag_square}
\end{figure}

\begin{figure}[tbh]
    \centering
    \includegraphics[width=1.\textwidth]{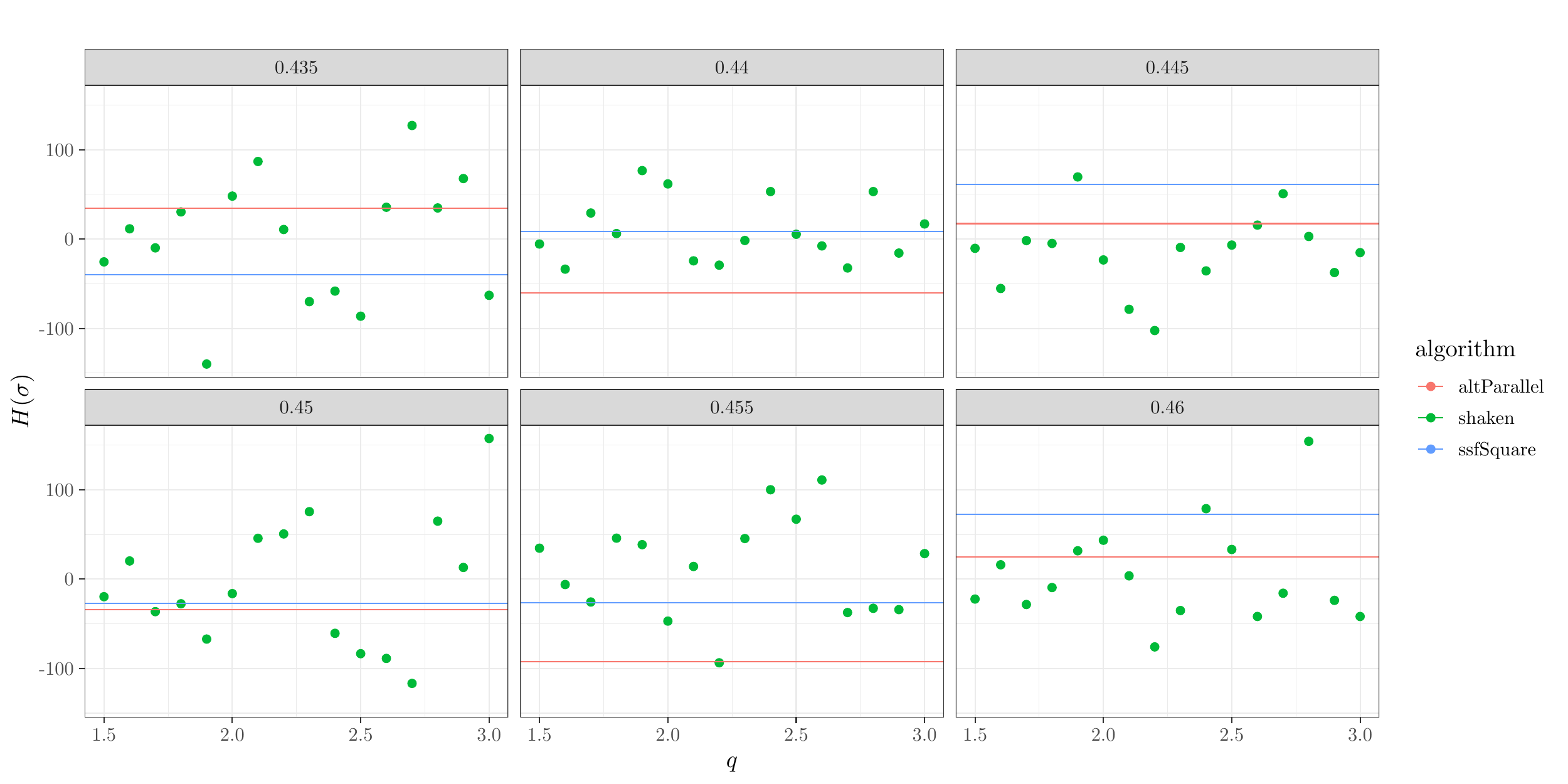}
    \caption{Sample average of the energy $H(\sigma)$ for several values of $J$}
    \label{fig:avg_H_square}
\end{figure}

\begin{figure}[tbh]
    \centering
    \includegraphics[width=1.\textwidth]{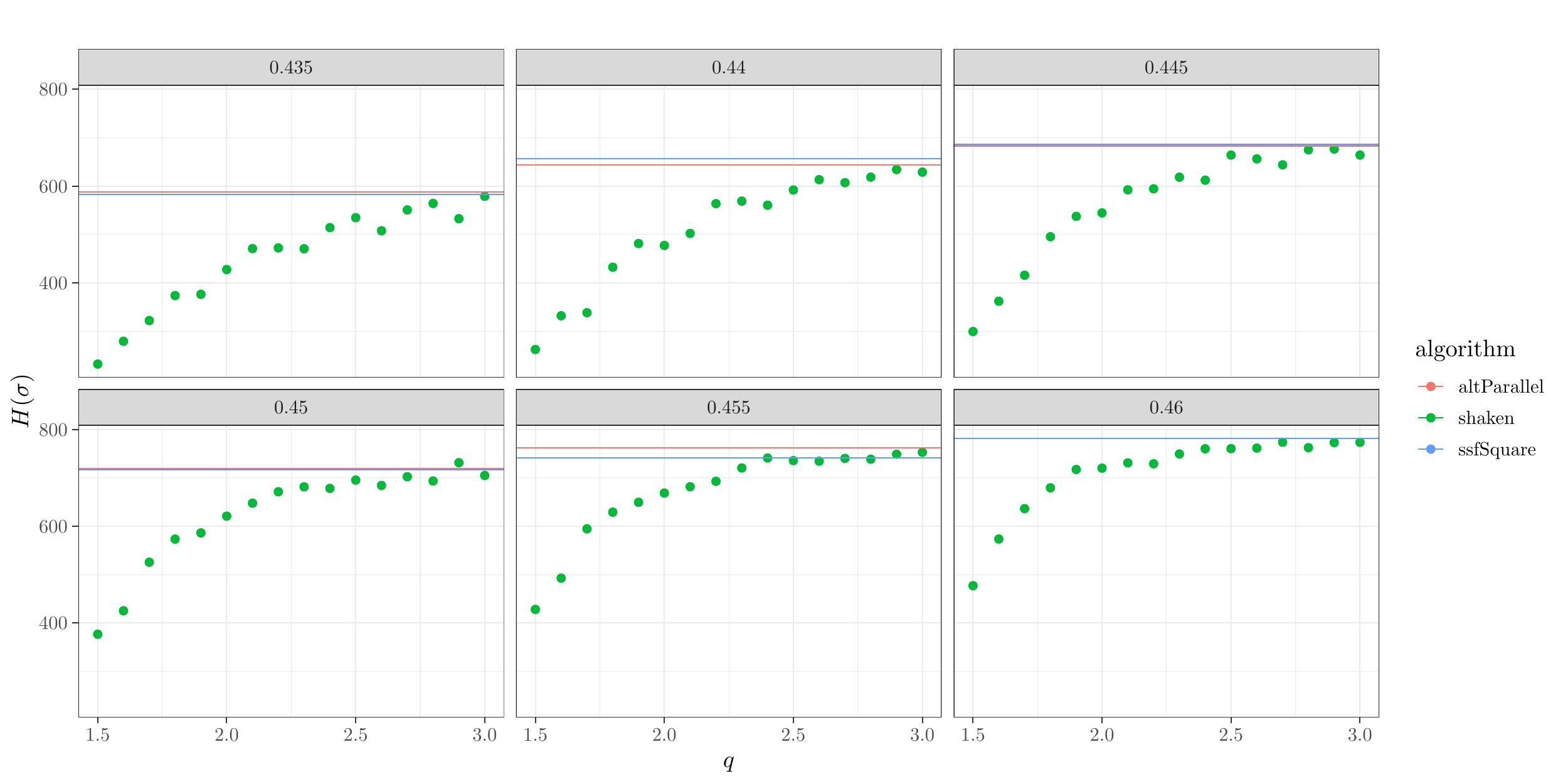}
    \caption{Sample standard deviation of the energy $H(\sigma)$ for several values of $J$}
    \label{fig:sd_H_square}
\end{figure}

In \cite{shakenDynamicsArxiv2019}, Theorem~2.3 it has been shown that, for large values of $q$, the equilibrium distribution of the shaken dynamics approaches the Gibbs measure for the Ising model on the square lattice. More precisely it has been proven that, if
\begin{align*}
    \lim_{|\Lambda| \to \infty} e^{-2q}|\Lambda| = 0,
\end{align*}
then, for $J$ sufficiently large,
\begin{align*}
    \lim_{|\Lambda| \to \infty} \tvar{\pi_s}{\pi_G} = 0,
\end{align*}
where $\pi_G$ is the Gibbs measure for the Ising model on the square lattice.
 Therefore it makes sense to evaluate numerically the goodness of this approximation as $q$ increases. To this purpose we consider two observable: the magnetization $m$ and the energy $H(\sigma)$. For both observable we compare their sample mean and sample standard deviation   over samples drawn from the equilibrium distribution of the shaken dynamics with the sample mean and the sample standard deviation of two other reference dynamics having the Gibbs measure as stationary distribution. One of the two reference dynamics taken into account is, again, the heat bath dynamics. The other dynamics is a parallel version of the heat bath dynamics that updates, alternatively, the spins on the odd and the even sites of the lattice. The latter is the equivalent for the square lattice of the alternate parallel dynamics on the hexagonal lattice defined by equations \ref{eq:leftTrans} and \ref{eq:rightTrans}. Theorem~2.2 in \cite{shakenDynamicsArxiv2019} states that the equilibrium measure of this dynamics is, indeed, the Gibbs measure on the square lattice. For all these dynamics, samples are drawn using the Propp-Wilson algorithm introduced above. Several values of $J$ close to the critical value for the Ising model on the square lattice and the results obtained are summarized in Figg.~\ref{fig:avg_mag_square}, \ref{fig:sd_mag_square}, \ref{fig:avg_H_square} and \ref{fig:sd_H_square}.

The data suggests that, for $q \ge 2.5$ the approximation provided by the shaken dynamics is quite good.

\begin{figure}[tbh]
    \centering
    \includegraphics[width=1.\textwidth]{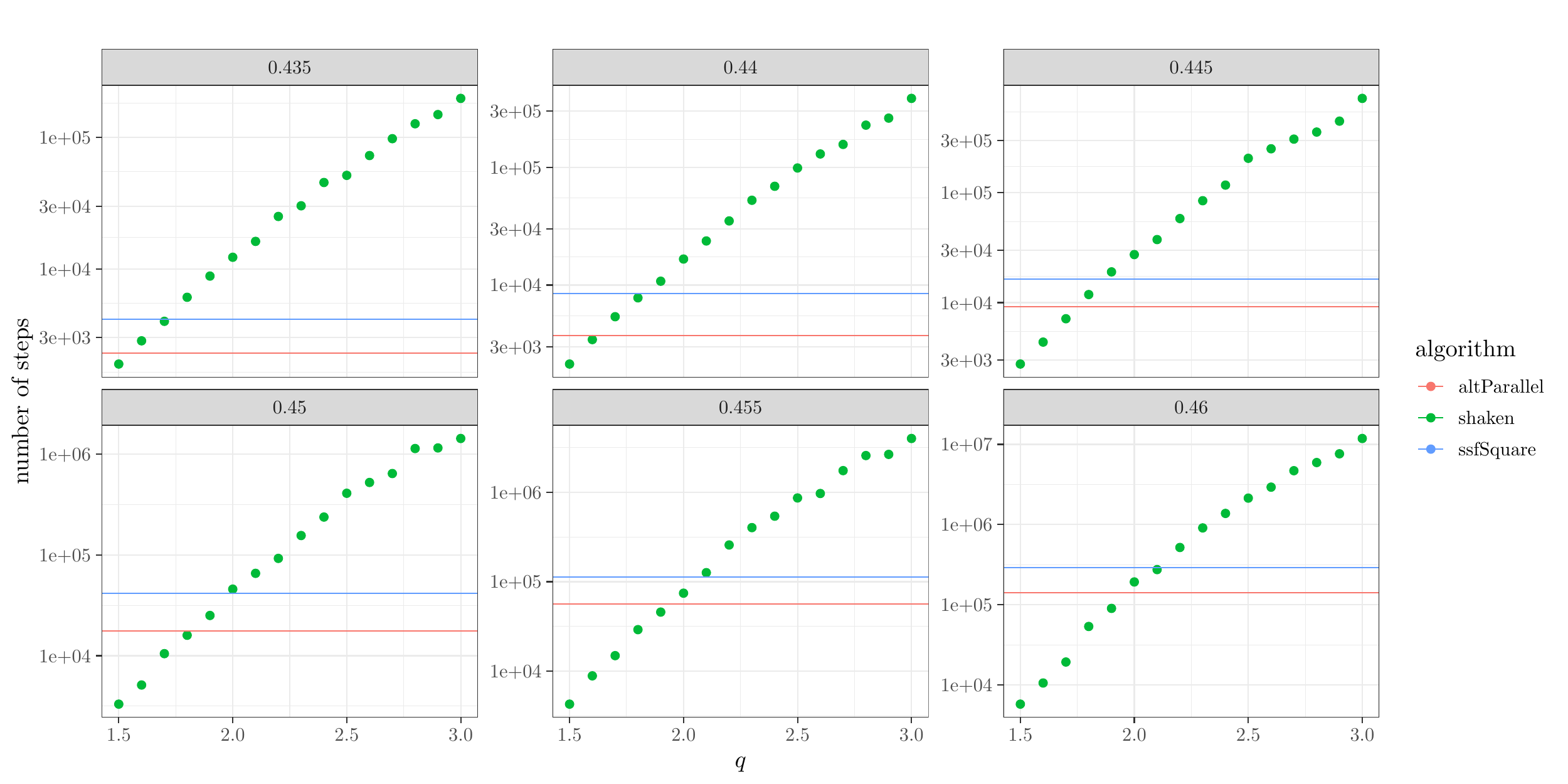}
    \caption{Sample average of the coalescence time (number of steps) for several values of $J$}
    \label{fig:avg_coal_time_square}
\end{figure}

On the other hand, we also estimated the time required to approach the equilibrium distribution by comparing the coalesce time of the shaken dynamics with those of the two other reference dynamics. Also in this case the number of steps required by the single spin flip dynamics is renormalized with the volume of the box $\Lambda$. The result obtained are summarized in Fig.~\ref{fig:avg_coal_time_square}. It is apparent that, though more flexible, the shaken dynamics becomes slower than ``specialized'' algorithms as the accuracy of the approximation increases.

\begin{figure}[h!]
\centering
\hfill
\subfigure[]
{\includegraphics[width=0.4\textwidth]{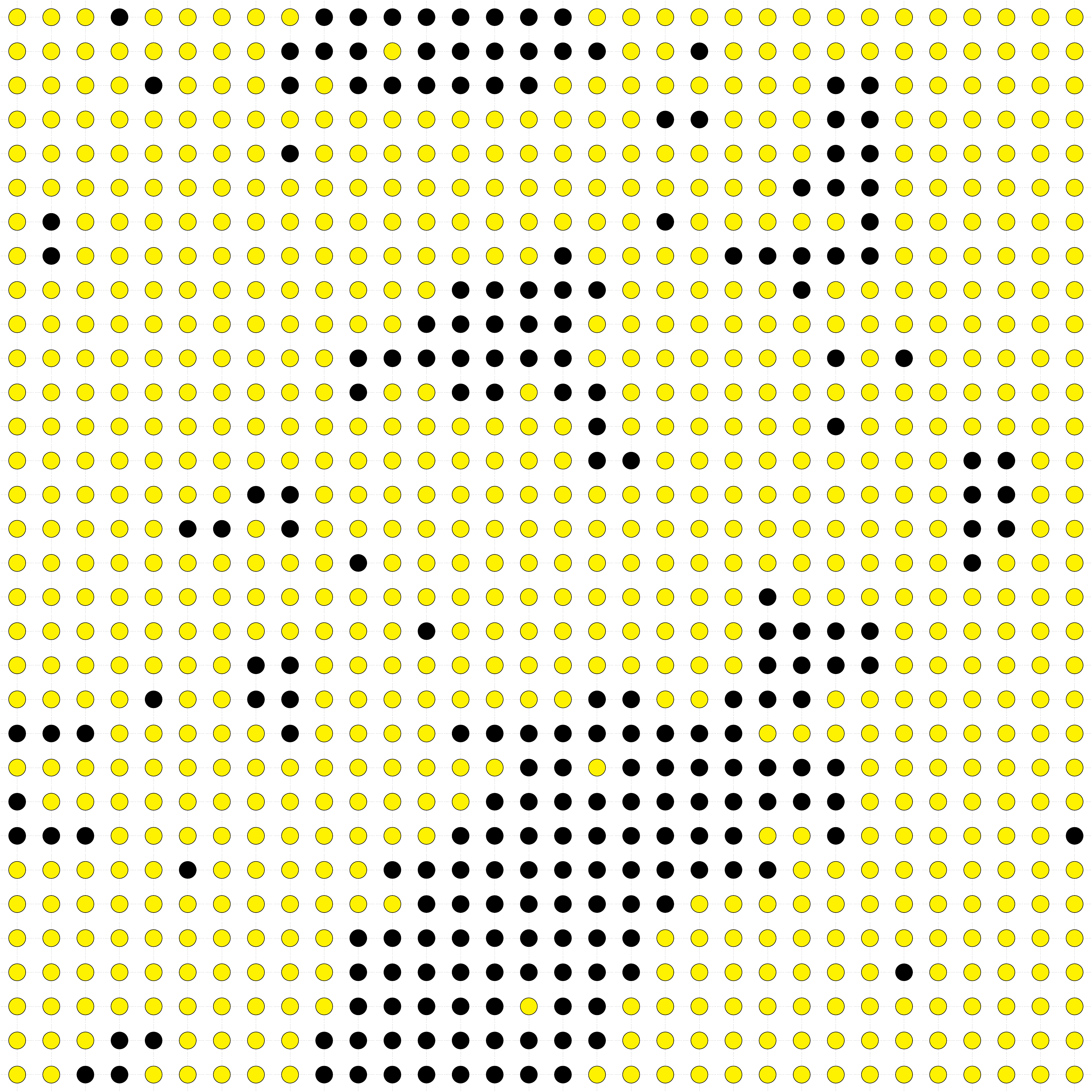}}
\hfill
\subfigure[]
{\includegraphics[width=0.4\textwidth]{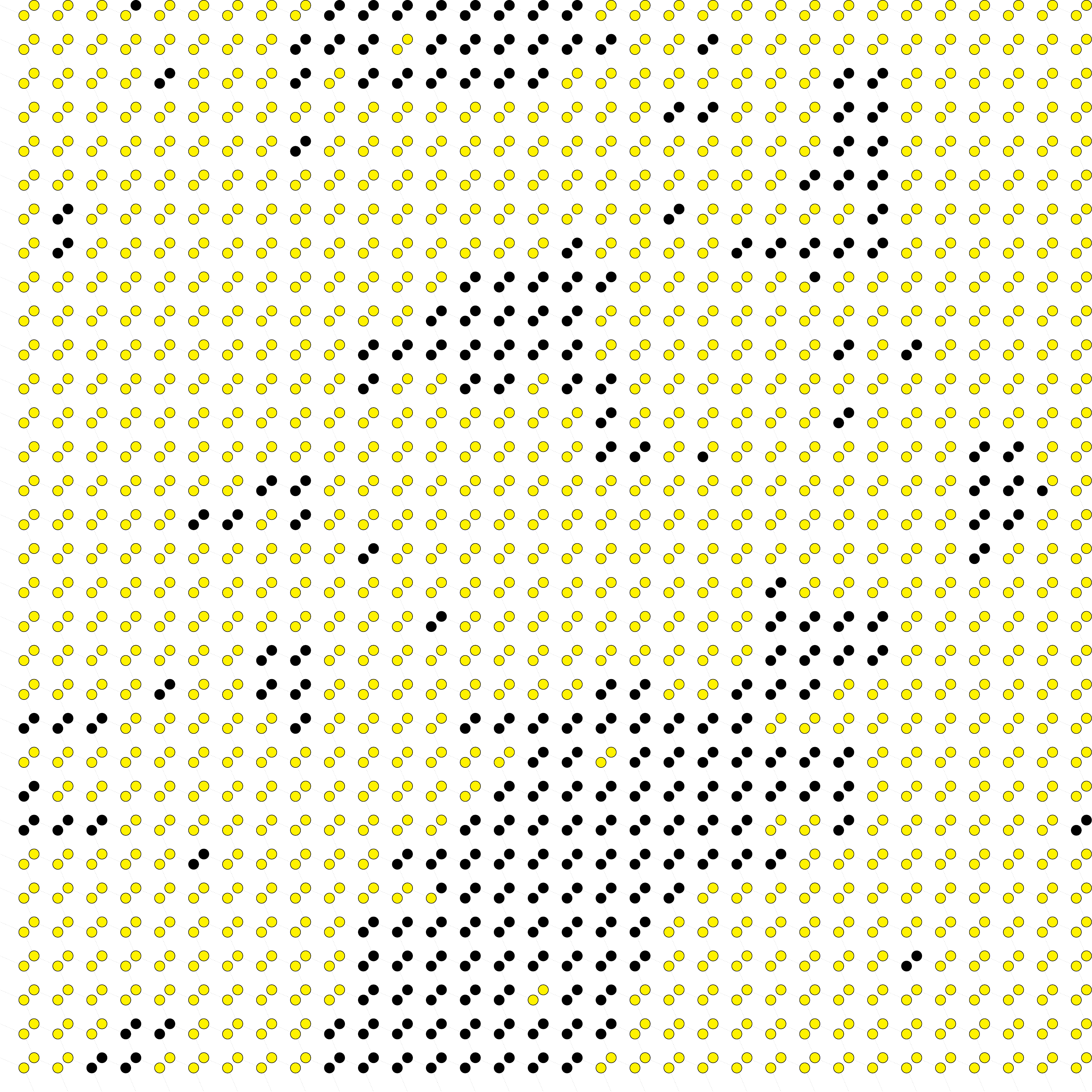}}
\hfill
\caption{$J = 0.44$, $q = 3.0$}
\label{fig:qLargeConfig}
\end{figure}

\begin{figure}[h!]
\centering
\hfill
\subfigure[]
{\includegraphics[width=0.4\textwidth]{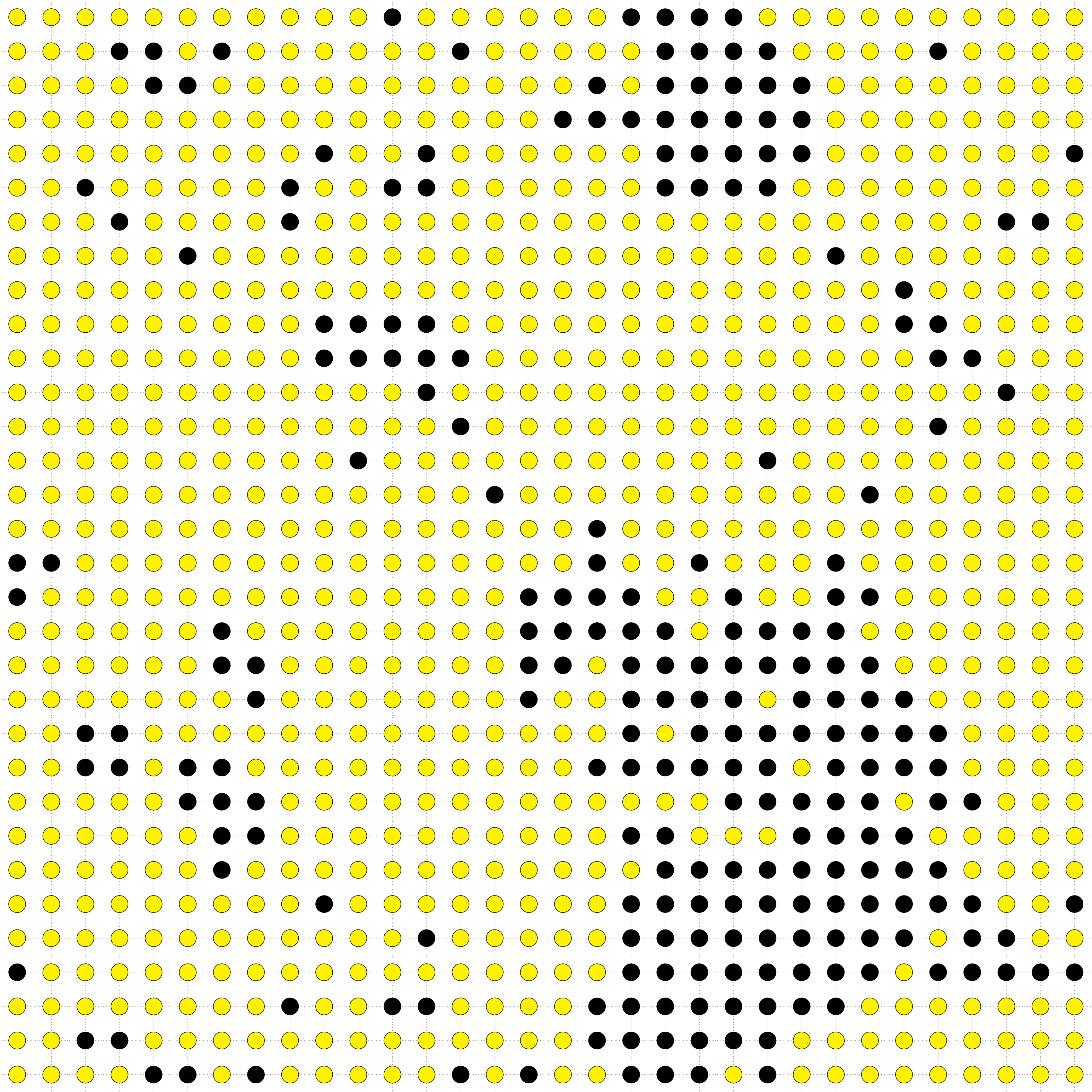}}
\hfill
\subfigure[]
{\includegraphics[width=0.4\textwidth]{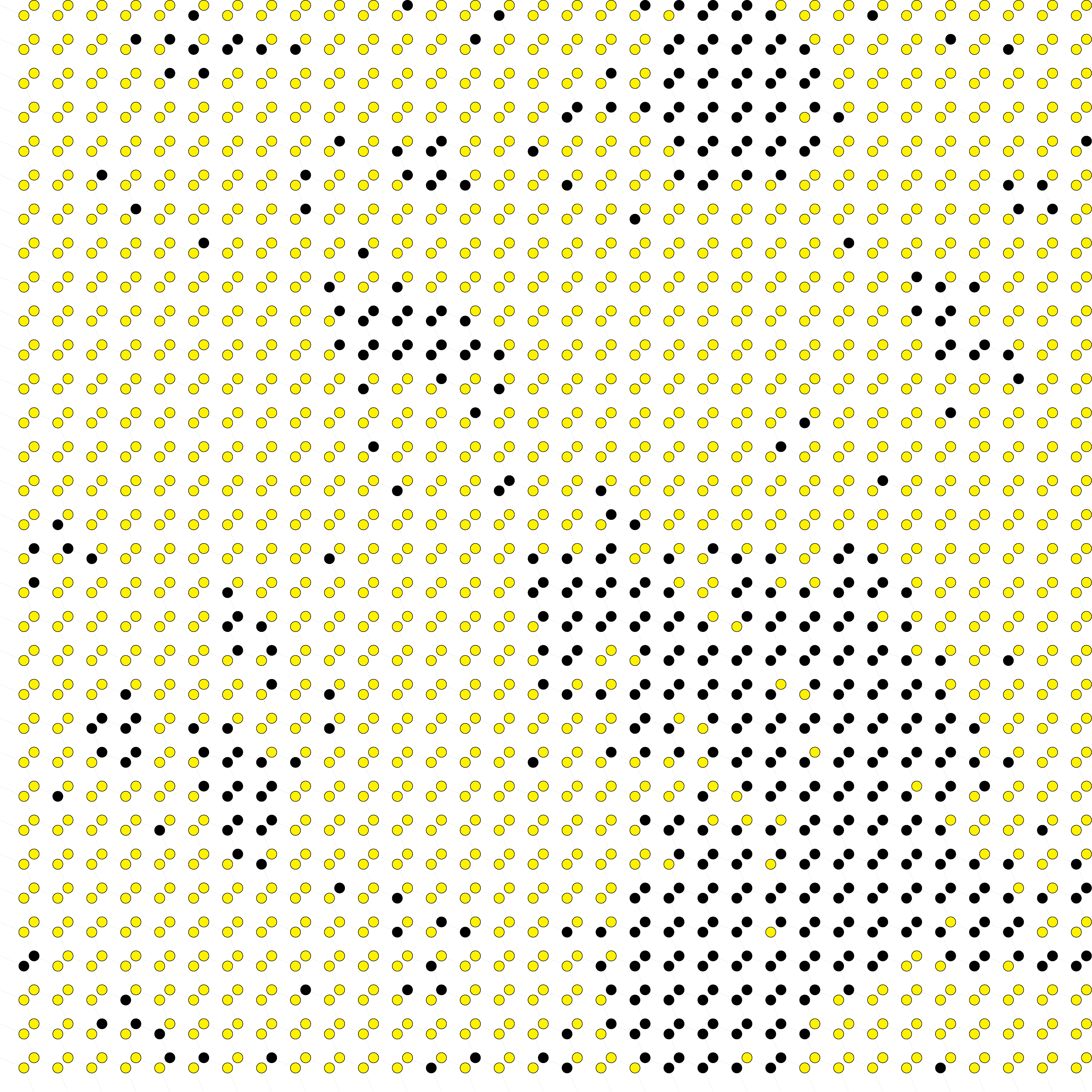}}
\hfill
\caption{$J = 0.6585$, $q = 0.6585$}
\label{fig:qBisectConfig}
\end{figure}

\begin{figure}[h!]
\centering
\hfill
\subfigure[]
{\includegraphics[width=0.4\textwidth]{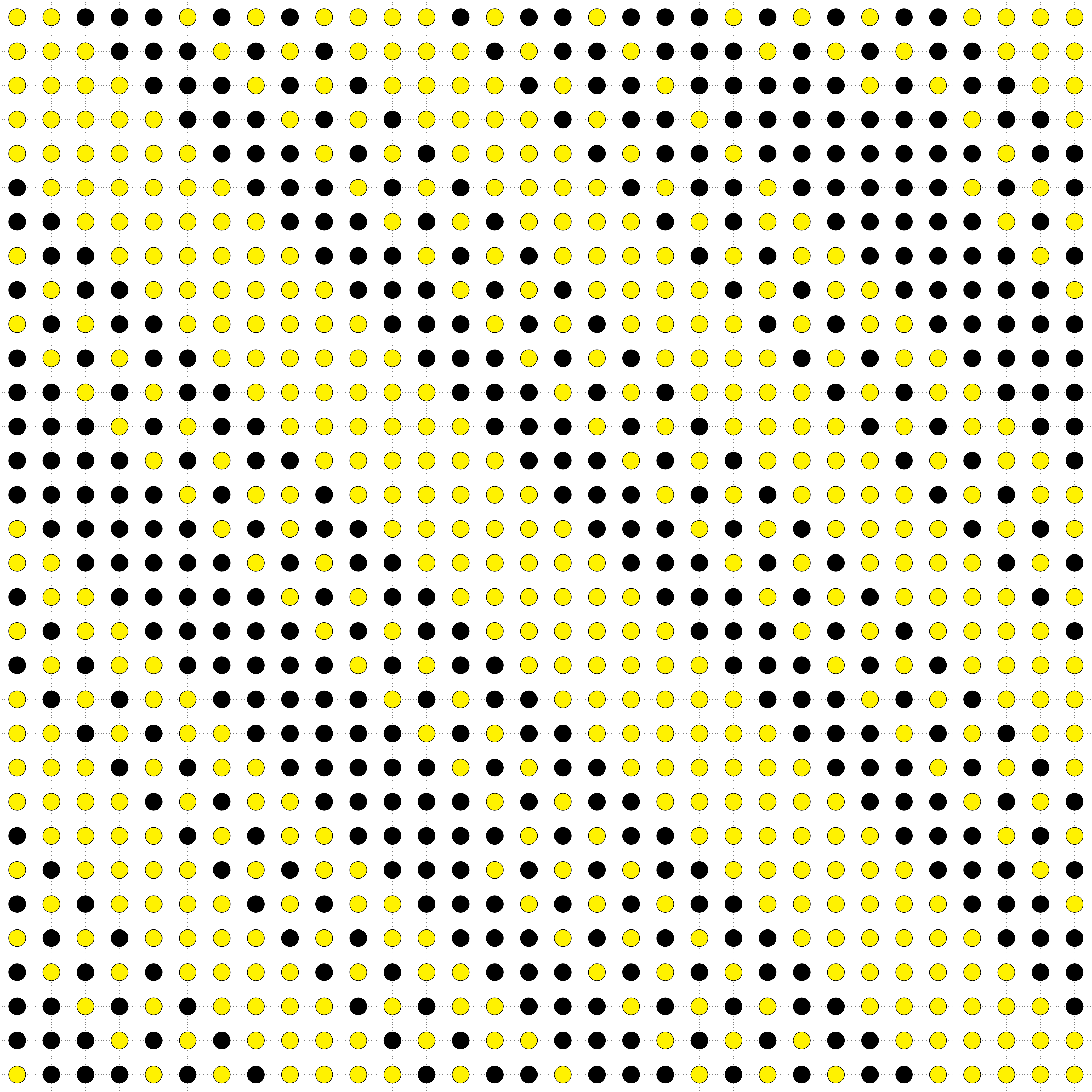}}
\hfill
\subfigure[]
{\includegraphics[width=0.4\textwidth]{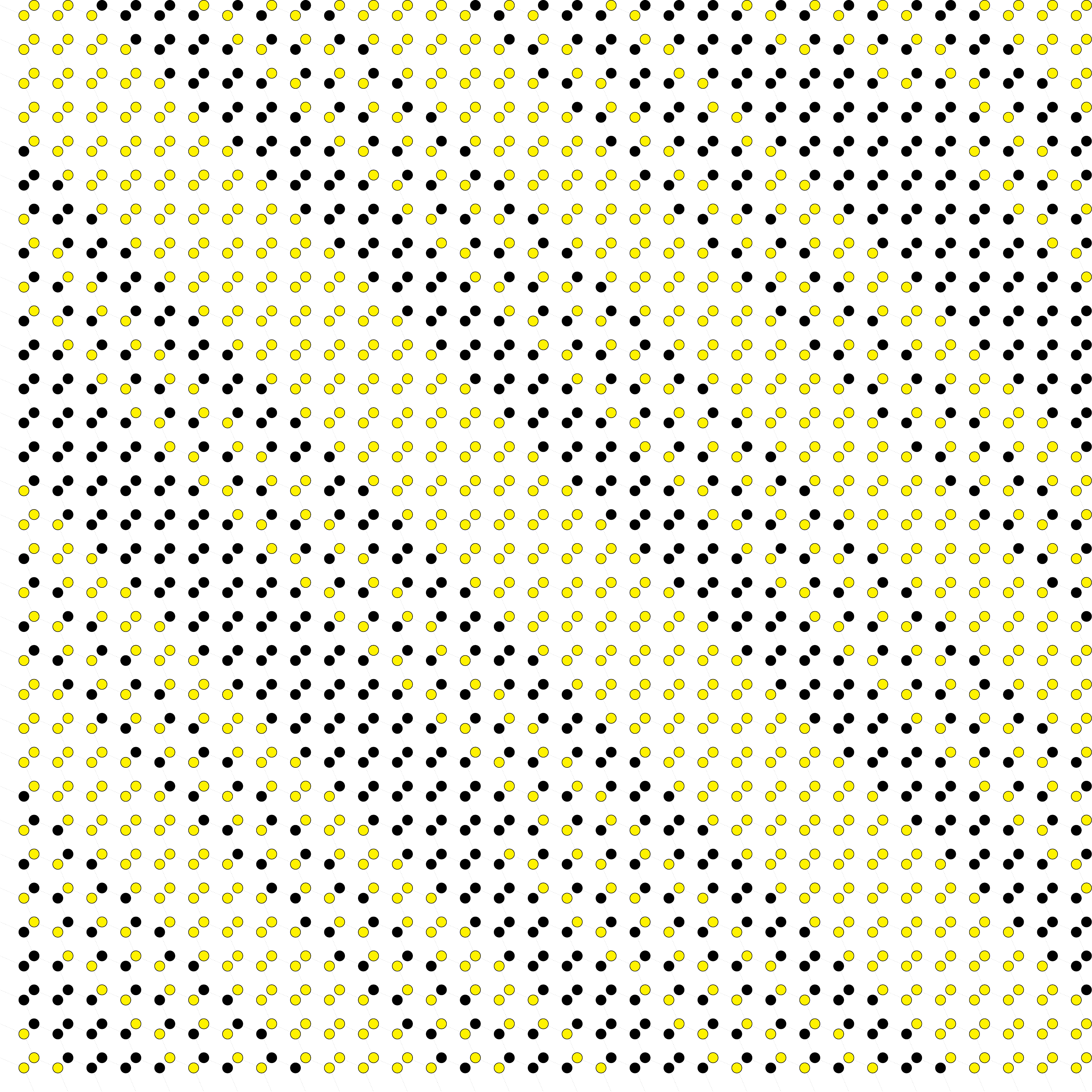}}
\hfill
\caption{$J = 2.0$, $q = 0.03$}
\label{fig:qSmallConfig}
\end{figure}

Parts (b) of Figg.~\ref{fig:qLargeConfig}, \ref{fig:qBisectConfig} and \ref{fig:qSmallConfig} show configurations drawn from the equilibrium distribution of the alternate parallel on the hexagonal whereas parts (a) show the corresponding sub-configurations on the sublattice $\Lambda_1$. These sub-configurations are, therefore, drawn from the equilibrium distribution of the shaken dynamics. In Fig.~\ref{fig:qLargeConfig} it is possible to observe that the spins linked by a $q$-edge have almost always the same value. This is in good accordance with the fact that stationary measure of the shaken dynamics is close to the Gibbs measure for the Ising model on the square lattice. On the other side, Fig.~\ref{fig:qSmallConfig} is consistent with the fact that for $q$ very small the equilibrium measure of the shaken dynamics tends to that of a colection of weakly dependent unidimensional Ising models.

\subsection{Correlations}

Theorem~2.4 in \cite{criticalitySquareToHexArxiv2019} establishes that, if $q$ is sufficiently small, $\pi(\sigma_{0,0}, \sigma_{\ell,\ell}) < \pi(\sigma_{0,\ell}, \sigma_{\ell,0})$ where $\pi$ is the equilibrium measure of the shaken dynamics and $\sigma$ is, therefore, a spin configuration living on $\Lambda_1$.
In words, the theorem states that the SW-NE correlations are weaker than the NW-SE ones if the \emph{self interaction} is weak. On the other hand, we expect that the SW-NE and the NW-SE correlations tend to be similar for large values of $q$, that is for those values of the pair $(q,J)$ for which the equilibrium distribution of the shaken dynamics approaches the Gibbs measure of the Ising model on the square lattice.

We study the SW-NE and the NW-SE correlations as $\ell$ varies with $\Lambda$ a $32 \times 32$ square box.
The results are shown in Table~\ref{tab:correlations}.


\thispagestyle{empty}
\begin{table}[]
\centering
\begin{tabular}{@{}llcllllll@{}}
\toprule
\multirow{2}{*}{$q$}   & \multirow{2}{*}{$J$}   & \multicolumn{1}{l}{\multirow{2}{*}{supercritical}} & \multirow{2}{*}{direction} & \multicolumn{5}{c}{$\ell$}                \\ \cmidrule(l){5-9}
                       &                        & \multicolumn{1}{l}{}                               &                            & 1     & 2      & 4      & 8      & 16     \\ \midrule
\multirow{2}{*}{0.05}  & \multirow{2}{*}{1.7}   & \multirow{2}{*}{}                                  & NW-SE                      & 0.821 & 0.765  & 0.7    & 0.481  & 0.425  \\ \cmidrule(l){4-9}
                       &                        &                                                    & SW-NE                      & 0.313 & -0.063 & -0.051 & 0.195  & 0.454  \\ \midrule
\multirow{2}{*}{0.05}  & \multirow{2}{*}{1.855} & \multirow{2}{*}{\checkmark}                         & NW-SE                      & 0.916 & 0.852  & 0.767  & 0.704  & 0.726  \\ \cmidrule(l){4-9}
                       &                        &                                                    & SW-NE                      & 0.618 & 0.316  & 0.124  & 0.081  & 0.739  \\ \midrule
\multirow{2}{*}{0.2}   & \multirow{2}{*}{1.05}  & \multirow{2}{*}{}                                  & NW-SE                      & 0.566 & 0.463  & 0.444  & 0.38   & -0.016 \\ \cmidrule(l){4-9}
                       &                        &                                                    & SW-NE                      & 0.4   & 0.203  & 0.002  & -0.037 & 0.041  \\ \midrule
\multirow{2}{*}{0.2}   & \multirow{2}{*}{1.175} & \multirow{2}{*}{\checkmark}                         & NW-SE                      & 0.84  & 0.7    & 0.624  & 0.584  & 0.54   \\ \cmidrule(l){4-9}
                       &                        &                                                    & SW-NE                      & 0.54  & 0.54   & 0.52   & 0.5    & 0.685  \\ \midrule
\multirow{2}{*}{0.4}   & \multirow{2}{*}{0.82}  & \multirow{2}{*}{}                                  & NW-SE                      & 0.65  & 0.507  & 0.462  & 0.356  & 0.426  \\ \cmidrule(l){4-9}
                       &                        &                                                    & SW-NE                      & 0.398 & 0.279  & 0.119  & 0.103  & 0.218  \\ \midrule
\multirow{2}{*}{0.4}   & \multirow{2}{*}{0.86}  & \multirow{2}{*}{\checkmark}                         & NW-SE                      & 0.68  & 0.644  & 0.541  & 0.679  & 0.538  \\ \cmidrule(l){4-9}
                       &                        &                                                    & SW-NE                      & 0.6   & 0.431  & 0.59   & 0.485  & 0.566  \\ \midrule
\multirow{2}{*}{0.6}   & \multirow{2}{*}{0.67}  & \multirow{2}{*}{}                                  & NW-SE                      & 0.74  & 0.646  & 0.4    & 0.378  & 0.167  \\ \cmidrule(l){4-9}
                       &                        &                                                    & SW-NE                      & 0.622 & 0.401  & 0.36   & 0.321  & 0.283  \\ \midrule
\multirow{2}{*}{0.6}   & \multirow{2}{*}{0.7}   & \multirow{2}{*}{\checkmark}                         & NW-SE                      & 0.855 & 0.772  & 0.763  & 0.732  & 0.664  \\ \cmidrule(l){4-9}
                       &                        &                                                    & SW-NE                      & 0.654 & 0.677  & 0.61   & 0.593  & 0.578  \\ \midrule
\multirow{2}{*}{0.65}  & \multirow{2}{*}{0.65}  & \multirow{2}{*}{}                                  & NW-SE                      & 0.701 & 0.472  & 0.477  & 0.503  & 0.475  \\ \cmidrule(l){4-9}
                       &                        &                                                    & SW-NE                      & 0.56  & 0.501  & 0.4    & 0.279  & 0.481  \\ \midrule
\multirow{2}{*}{0.663} & \multirow{2}{*}{0.663} & \multirow{2}{*}{\checkmark}                         & NW-SE                      & 0.749 & 0.646  & 0.6    & 0.544  & 0.477  \\ \cmidrule(l){4-9}
                       &                        &                                                    & SW-NE                      & 0.65  & 0.52   & 0.442  & 0.578  & 0.642  \\ \midrule
\multirow{2}{*}{0.8}   & \multirow{2}{*}{0.58}  & \multirow{2}{*}{}                                  & NW-SE                      & 0.68  & 0.627  & 0.281  & 0.243  & 0.245  \\ \cmidrule(l){4-9}
                       &                        &                                                    & SW-NE                      & 0.609 & 0.522  & 0.307  & 0.444  & 0.433  \\ \midrule
\multirow{2}{*}{0.8}   & \multirow{2}{*}{0.61}  & \multirow{2}{*}{\checkmark}                         & NW-SE                      & 0.66  & 0.661  & 0.62   & 0.581  & 0.52   \\ \cmidrule(l){4-9}
                       &                        &                                                    & SW-NE                      & 0.74  & 0.52   & 0.581  & 0.52   & 0.524  \\ \midrule
\multirow{2}{*}{1.0}   & \multirow{2}{*}{0.52}  & \multirow{2}{*}{}                                  & NW-SE                      & 0.581 & 0.56   & 0.258  & -0.019 & 0.103  \\ \cmidrule(l){4-9}
                       &                        &                                                    & SW-NE                      & 0.541 & 0.299  & 0.341  & 0.221  & 0.04   \\ \midrule
\multirow{2}{*}{1.0}   & \multirow{2}{*}{0.55}  & \multirow{2}{*}{\checkmark}                         & NW-SE                      & 0.602 & 0.606  & 0.398  & 0.441  & 0.599  \\ \cmidrule(l){4-9}
                       &                        &                                                    & SW-NE                      & 0.58  & 0.58   & 0.561  & 0.54   & 0.532  \\ \midrule
\multirow{2}{*}{2.5}   & \multirow{2}{*}{0.43}  & \multirow{2}{*}{}                                  & NW-SE                      & 0.462 & 0.456  & 0.27   & 0.194  & 0.164  \\ \cmidrule(l){4-9}
                       &                        &                                                    & SW-NE                      & 0.541 & 0.42   & 0.221  & 0.26   & 0.201  \\ \midrule
\multirow{2}{*}{2.5}   & \multirow{2}{*}{0.46}  & \multirow{2}{*}{\checkmark}                         & NW-SE                      & 0.658 & 0.701  & 0.74   & 0.701  & 0.699  \\ \cmidrule(l){4-9}
                       &                        &                                                    & SW-NE                      & 0.761 & 0.739  & 0.654  & 0.538  & 0.654  \\ \bottomrule
\end{tabular}
\caption{Spin-spin correlations. The checkmark \checkmark in the \emph{supercritical} column identifies pairs $(q,J)$ above the critical curve $J_c$}
\label{tab:correlations}
\end{table}

All pairs $(q, J)$ taken into account correspond to points of the $q, J$ plane close to the critical curve $J_c(q)$. It is possible to observe that, as $q$ decreases, the SW-NE correlations become, indeed, smaller than the NW-SE ones, whereas, for $q$ large the two are quite similar. Further, if the pair $(q,J)$ is below the critical curve the correlations decay quite rapidly. On the other hand, if $(q, J)$ is above $J_c$ the correlations are significant also for larger values of $\ell$.

\section{Implementation details}\label{sec:implementation_details}

To approximate numerically the critical curve $J_c(q)$, we take samples for different values of $J$ and $q$. The code used for the simulationin written in Julia \cite{JuliaLanguage} and simulations are performed  through 80 thread processors running, in parallel, the simulation on 80 couples of values $(q,J)$ in the range of $(q,J) \in (0,2)\times (0,2)$.
The Hamiltonian is defined on a square $200\times 200$ lattice.
Statistics are collected over  300,000 iterations.
Fig.\ref{fig:niagara} shows that the chosen simulation parameter is good enough to approximate the critical curve.

The elementary step of the shaken dynamics described in the previous section has been simulated by the
Algorithm~\ref{algo:shaken_dynamics}.
A spin configuration is updated via a sequence of two similar half steps.
The computation of the vector of local fields $h$ that drives the transition probabilities of each spin is alternatively carried out
using the functions {\tt collectUR} and {\tt collectDL} which determine the {\tt up-right} and {\tt down-left} contribution as in eq. (\ref{eq:urdl}).
\begin{algorithm}
	\caption{collectUR}
	\label{algo:clUR}
	\begin{algorithmic}[1]
		\REQUIRE $x_{\sigma},J,q$
		\ENSURE $f$
		\STATE $f \gets J(\spinUp{x}{\sigma} + \spinRight{x}{\sigma}) + q \spin{x}{\sigma}$
		\STATE Return $f$
	\end{algorithmic}
\end{algorithm}
\begin{algorithm}
	\caption{collectDL}
	\label{algo:clDL}
	\begin{algorithmic}[1]
		\REQUIRE $x_{\sigma},J,q$
		\ENSURE $f$
		\STATE $f \gets J(\spinDown{x}{\sigma} + \spinLeft{x}{\sigma}) + q \spin{x}{\sigma}$
		\STATE Return $f$
	\end{algorithmic}
\end{algorithm}\hfill \break
The algorithm \ref{algo:shaken_dynamics} is the complete update in two steps of the shaken dynamics, which is more general than the one used in \cite{Lancia2013}.
\begin{algorithm}[!h]
	\caption{Shaken dynamics}
	\label{algo:shaken_dynamics}
	\begin{algorithmic}[1]
		\REQUIRE initial spin configuration $\sigma$
		\ENSURE updated spin configuration $\tau$
		\FOR{each $x_{\sigma}$}
		\STATE $h \gets$ \textbf{collectUR}($x_{\sigma},J,q$)
		\STATE $p \gets exp(h)/2\cosh(h)$
		\IF{$ rand() < p$}
		\STATE $x_{\tau} \gets 1$
		\ELSE
		\STATE $x_{\tau} \gets -1$
		\ENDIF
		\ENDFOR
		\STATE $\sigma \gets \tau$
		\FOR{each $x_{\sigma}$}
		\STATE $h \gets$ \textbf{collectDL}($x_{\sigma},J,q$)
		\STATE $p \gets exp(h)/2\cosh(h)$
		\IF{$ rand() < p$}
		\STATE $x_{\tau} \gets 1$
		\ELSE
		\STATE $x_{\tau} \gets -1$
		\ENDIF
		\ENDFOR
	\end{algorithmic}
\end{algorithm}

The choice of collecting the statistics over 300,000 time steps (after a warm up time of 300,000 additional stime stesps) turned out to be good enough, and the results show, unmistakably,the separation of the two phases (ordered and disordered).

We implemented the algorithm \ref{algo:shaken_dynamics} in two parallel ways.
A \textbf{CUDA} \footnote{Compute Unified Device Architecture, parallel platform and programming model to make use of the Graphic Processing Units general purpose computing simple and elegant.} implementation of a parallel heat bath for large dimension Lattice spin, and a \textbf{Julia} \cite{JuliaLanguage} implementation on a single CPU to be used on a multiprocessor systems (trivial parallel on a multi data input). Both have been optimized to handle our problem, and used to simulate the shaken dynamics of the PCA, a quasi-similar behavior was observed during our experiment.
\subsubsection{Parallel single-GPU code}

The general heat bath procedure has been implemented on SIMD (Single Instruction Multiple Data) system.
To optimize the code exploiting the CUDA memory architecture we implemented three kernels for the functions {\tt collectUR}, {\tt collectDL} and for updating the configuration.
We used the default random generator from curand library.

The {\tt collect} function computes the transition probabilities in the given direction.
Each thread handles one spin on the lattice field.
The principal use of the global memory is the four square spin lattice, the two configuration sigma ($\sigma$) and tau ($\tau$), the fields which handles the Hamiltonian computation and the random-unit contains random uniform variables.

In our implementation all the operation are performed on register cache memory.
We did not use shared memory for the random-unit.
The code was written to run on the Nvidia-GPU Tesla P100 using by 16GB video memory, using 4 matrices of dimension $L\times L$, two for the lattice spin field (single byte), and two for the collected fields and for the random uniform (four byte).
All the matrices are allocated on the global memory.

For the management of the memory, before allocating the memory of the 4 matrices, the code used approximately 303 MB, leaving $15973.250000$ MB.
We used $2*4*L*L + 2*L*L$ bytes, but we can not go beyond $10^5$ for this GPU.\\
The purpose of the CUDA implementation is to work on large dimension which allows us to observe the statistical behavior of the shaken dynamic, also for real time simulation.

\begin{figure}[h!]
	\centering
	\includegraphics[width=0.8\linewidth]{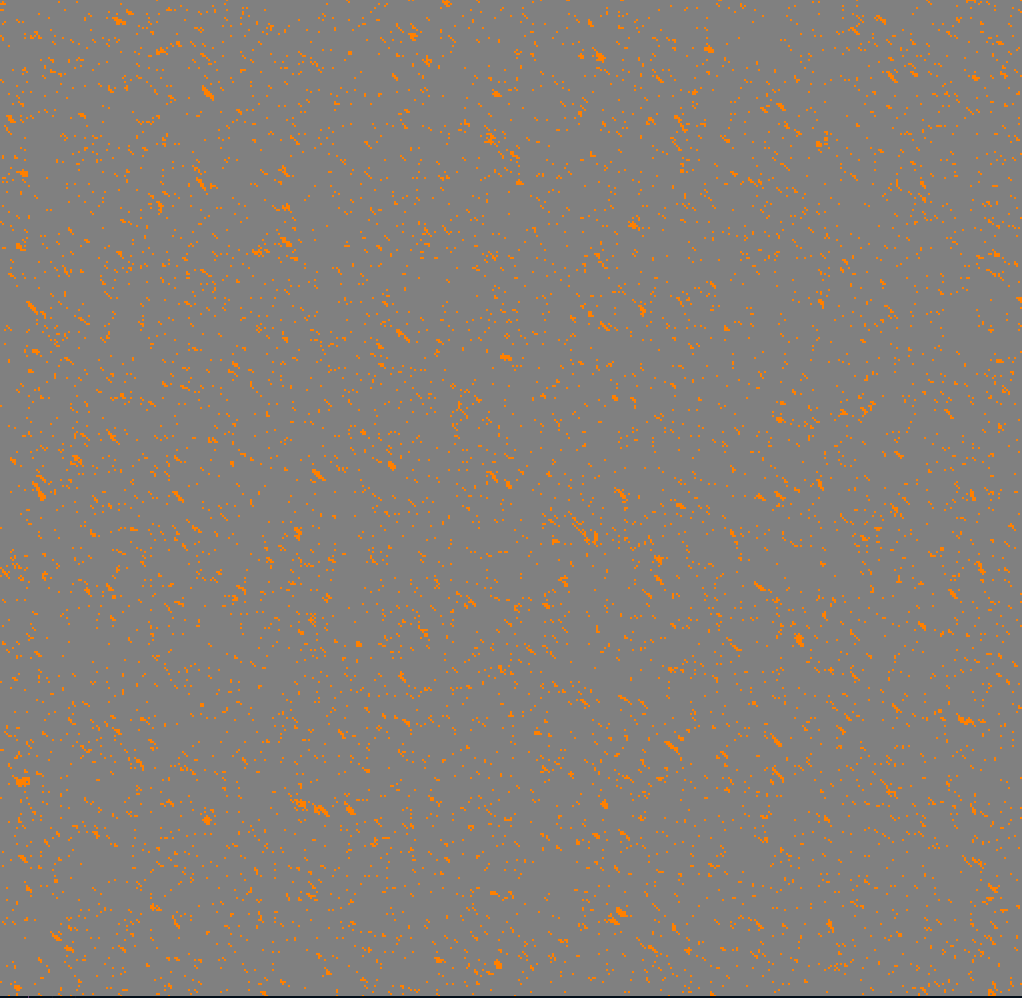}
	\caption{GPU sample: $L=512$, $J=0.99$, $q=0.5$, iteration$=60^{th}$}
	\label{fig:exampl1}
\end{figure}

Fig. (\ref{fig:exampl1}) shows a state of configuration captured at 60-th iteration on a simulation of the shaken dynamics for PCA lattice square with dimension $512\times512$, under the temperature $J = 0.99$ and the external field $q=0.5$.

\subsubsection{Benchmarking}
\label{subsub:benchmark}
\begin{figure}[h!]
	\centering
	\includegraphics[width=0.8\linewidth]{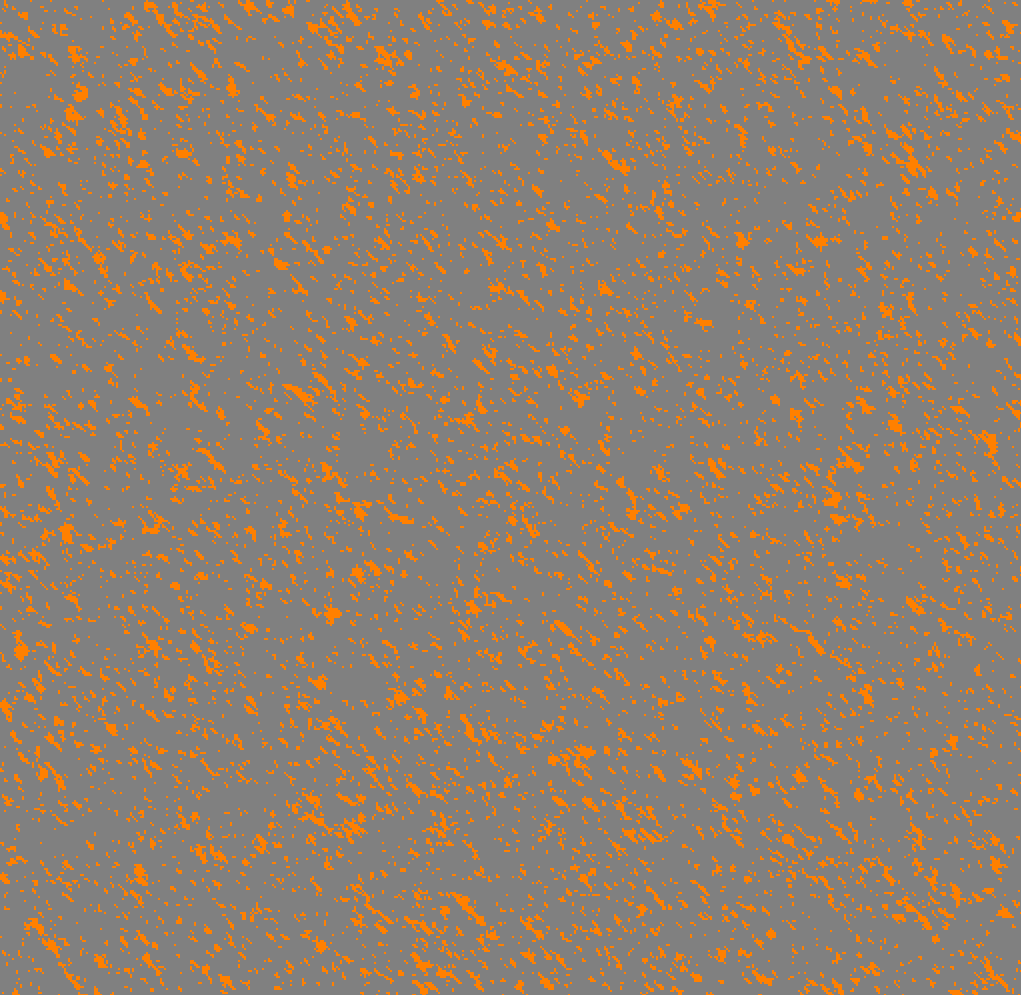}
	\caption{CPU sample: $L=512$, $J=0.99$, $q=0.5$, iteration$=60^{th}$}
	\label{fig:exampl2}
\end{figure}

To measure the performance of our GPU code it is not fair to compare it with the single-CPU implementation from Julia. We have implemented a serial version of the shaken dynamics in a lower level language, a captured sample for a square lattice spins of size $512\times512$ with $J=0.99$ and $q=0.5$ is given in Fig. (\ref{fig:exampl2}).
 For our simple measurement, we set the parameters $J=0.44$ and $q=0.66$ and to have more significant value we measure it in milliseconds. We compare the two implementation for different dimension $L$, hence the size of the square lattice which is $L^{2}$.
\begin{figure}[h!]
	\begin{tikzpicture}
	\begin{loglogaxis}[height=10cm, width=14cm, xlabel={Dimension $L$}, ylabel={Time in milliseconds}]
	\addplot[mark=x, color=gray]  table [x=size, y=time, col sep=comma] {Figure/cputime.csv};
	\addlegendentry{single-CPU}

	\addplot[mark=x, color=orange]  table [x=size, y=time, col sep=comma] {Figure/gputime.csv};
	\addlegendentry{single-GPU}

	\end{loglogaxis}
	\end{tikzpicture}
	\caption{Running-time in function of the size $L$.}
	\label{fig:bench}
\end{figure}
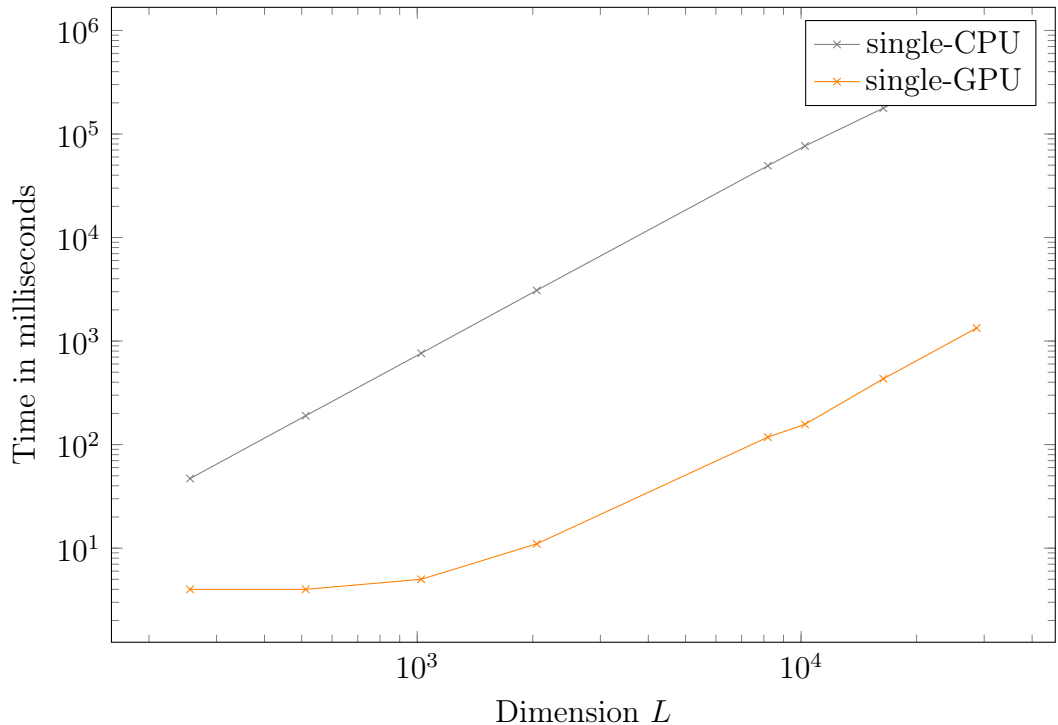
For this benchmark (Fig. (\ref{fig:bench})), we used an Nvidia graphic card Tesla P-100 vs single core of the CPU Intel(R) Xeon(R) CPU E5-2698 v4 @ 2.20GHz. We measure the time for one update execution. As we have mentioned the GPU memory is limited so that we did the experiment under this condition for the size of the square lattice spin.

We observe that the GPU is much faster than the CPU with a factor of 500 as far as the lattice size grows. We can see that the CPU time looks linear while the case for GPU is when the size is more than $2048\times2048$.

\section{Conclusions}
\label{par4}
\subsection{Summary}
The present work is a numerical experiment on the 2D Ising model. In particular the tasks are mainly focused on the {\sl shaken dynamics}, in which we used to approximate numerically the critical curve which relate the parameter $J$ and $q$, it separates the two phases on the region of parameters $(J,q)$. Also we are able to evaluate numerically the fact that the equilibrium distribution of this dynamics are close to Gibbs measure on the region when $q$ is large \cite{shakenDynamicsArxiv2019}. We provide two different parallel implementation perspective of the algorithm in which we give a benchmark to identify a speed-up that we can gain on using GPU.

As a MCMC algorithm we also give a comparison on the convergence to the equilibrium state of the algorithm by means of coalescence time. In this purpose we compare the coalescence time between the alternate and shaken dynamics on the critical line (red line in Fig. \ref{Jqplot}), this is followed by further discussion on the numerical aspect of the algorithm.

\subsection{Work in progress}
Most of the implementations we have used in this project are in Julia \cite{JuliaLanguage}. As pointed out above, the paper is a numerical supports for the two papers \cite{shakenDynamicsArxiv2019} and \cite{criticalitySquareToHexArxiv2019}. The code is quite complete for an academic use on the simulation of 2D Ising model, especially it contains the class of all the dynamics we have studied in this project and their methods. Our attempt is to provide a Julia library that can be used as framework to study the dynamics of the planar Ising model.

\section{Acknowledgements}
\label{sec:ackno}
The authors are thankful to Valentina Apollonio, Benedetto Scopppola and Elisabetta Scoppola
for their support and their precious suggestions.
A.T. has been supported by Project FARE 2016 Grant R16TZYMEHN

\printbibliography

\end{document}